\documentclass[]{article}

\usepackage{graphicx}
\usepackage{subfig}
\usepackage{amsmath}
\usepackage{amsfonts}
\usepackage[ruled,vlined,linesnumbered]{algorithm2e}
\graphicspath{{figs/}}

\newlength\mylen

\begin{document}

\title{Meshing strategies for the alleviation of mesh-induced effects in cohesive element models}

\author{J.~J.~Rimoli
\footnote{Correspondence to: Julian J. Rimoli, School of Aerospace Engineering, Georgia Institute of Technology, 270 Ferst Drive, Atlanta, GA 30332-0150, U.S.A.}
$^,$\footnote{E-mail: rimoli@gatech.edu} 
 and J.~J.~Rojas\\
School of Aerospace Engineering,\\
Georgia Institute of Technology,
Atlanta, GA 30332}

\maketitle

\begin{abstract}
One of the main approaches for modeling fracture and crack propagation in solid materials is adaptive insertion of cohesive elements, in which line-like (2D) or surface-like (3D) elements are inserted into the finite element mesh to model the nucleation and propagation of failure surfaces. In this approach, however, cracks are forced to propagate along element boundaries, following paths that in general require more energy per unit crack extension (greater driving forces) than those followed in the original continuum, which in turn leads to erroneous solutions. In this work we illustrate how the introduction of a discretization produces two undesired effects, which we term mesh-induced anisotropy and mesh-induced toughness. Subsequently, we analyze those effects through polar plots of the path deviation ratio (a measure of the ability of a mesh to represent straight lines) for commonly adopted meshes. Finally, we propose to reduce those effects through K-means meshes and through a new type of mesh, which we term conjugate-directions mesh. The behavior of all meshes under consideration as the mesh size is reduced is analyzed through a numerical study of convergence.
\end{abstract}

\section{Introduction}

The classical cohesive theory of fracture finds its origins in the pioneering works by Dugdale, Barenblatt and Rice \cite{dugdale1960jmps, barenblatt1962aam, rice1968fat}. In their work,  fracture is regarded as a progressive phenomenon in which crack formation takes place across a cohesive zone ahead of the crack tip and is resisted by cohesive tractions. Cohesive zone models are widely adopted by scientists and engineers perhaps due to their straightforward implementation within the traditional finite element formulation. Some of the mainstream technologies proposed to introduce the cohesive theory of fracture into finite element analysis are the eXtended Finite Element Method (X-FEM) and cohesive elements.

Sukumar et al. \cite{sukumar2000ijnme} first utilized the X-FEM for modeling 3D crack growth by adding a discontinuous function and the asymptotic crack tip field to the finite elements. Subsequently, the method was extended to account for cohesive cracks \cite{moes2002efm}. It is worth noting that while the X-FEM approach can potentially deal with arbitrary crack paths, it becomes increasingly complicated for problems involving pervasive fracture and fragmentation.

On the other hand, the cohesive element approach consists on the insertion of cohesive finite elements along the edges or faces of the 2D or 3D mesh correspondingly  \cite{xu1994jmps,xu1995ijf,camacho1996ijss,ortiz1999ijnme}. Even though this approach is well suited for problems involving pre-defined crack directions, a number of known issues affect the its accuracy when dealing with simulations including arbitrary crack paths, namely, (i) problems with the propagation of elastic stress waves (artificial compliance), (ii) spurious crack tip speed effects (lift-off), and (iii) mesh dependent effects (c.f. \cite{seagraves2009dfms} for a comprehensive review). Despite these well known limitations, the robustness of the method makes it one of the most common approaches for pervasive fracture and fragmentation analysis.

Some of the limitations present in early approaches to cohesive element models were addressed by subsequent research efforts. For example, artificial compliance and lift-off effects can be avoided by using an initially rigid cohesive law \cite{camacho1996ijss} or, more elegantly, a discontinuous Galerkin formulation with an activation criterion for cohesive elements \cite{noels2008ijnme, radovitzky2011cmame}. However, the problem of mesh dependency is still an active area of research.

Mesh-dependent effects are direct consequence of cracks being able to propagate only across boundaries between bulk finite elements. That is, the topology of the mesh forces cracks to follow paths that in general require more energy per unit crack extension (greater driving forces) than those followed in the original continuum. In this work, we first focus on the effects that common mesh topologies have on two main mesh-dependent effects, namely: mesh-induced toughness and mesh-induced anisotropy. We then illustrate how to decrease mesh-induced anisotropy through K-means meshes. Finally, we introduce a new type of mesh, termed conjugate-directions mesh, which greatly alleviate both effects.

The reminder of the paper is organized as follows: in section \ref{sec:MIE} we introduce a few preliminary concepts and discuss the mesh dependent effects; in section \ref{sec:CDM}, we define K-means and conjugate-directions meshes and we study their mesh-dependent behavior, including a numerical convergence analysis; finally, in section \ref{sec:SUM}, we summarize the main conclusions of the article.

\section{Mesh-induced effects}\label{sec:MIE}

\subsection{Basic definitions}
\begin{figure}
\begin{center}
\subfloat[]{\includegraphics[height=0.3\textwidth]{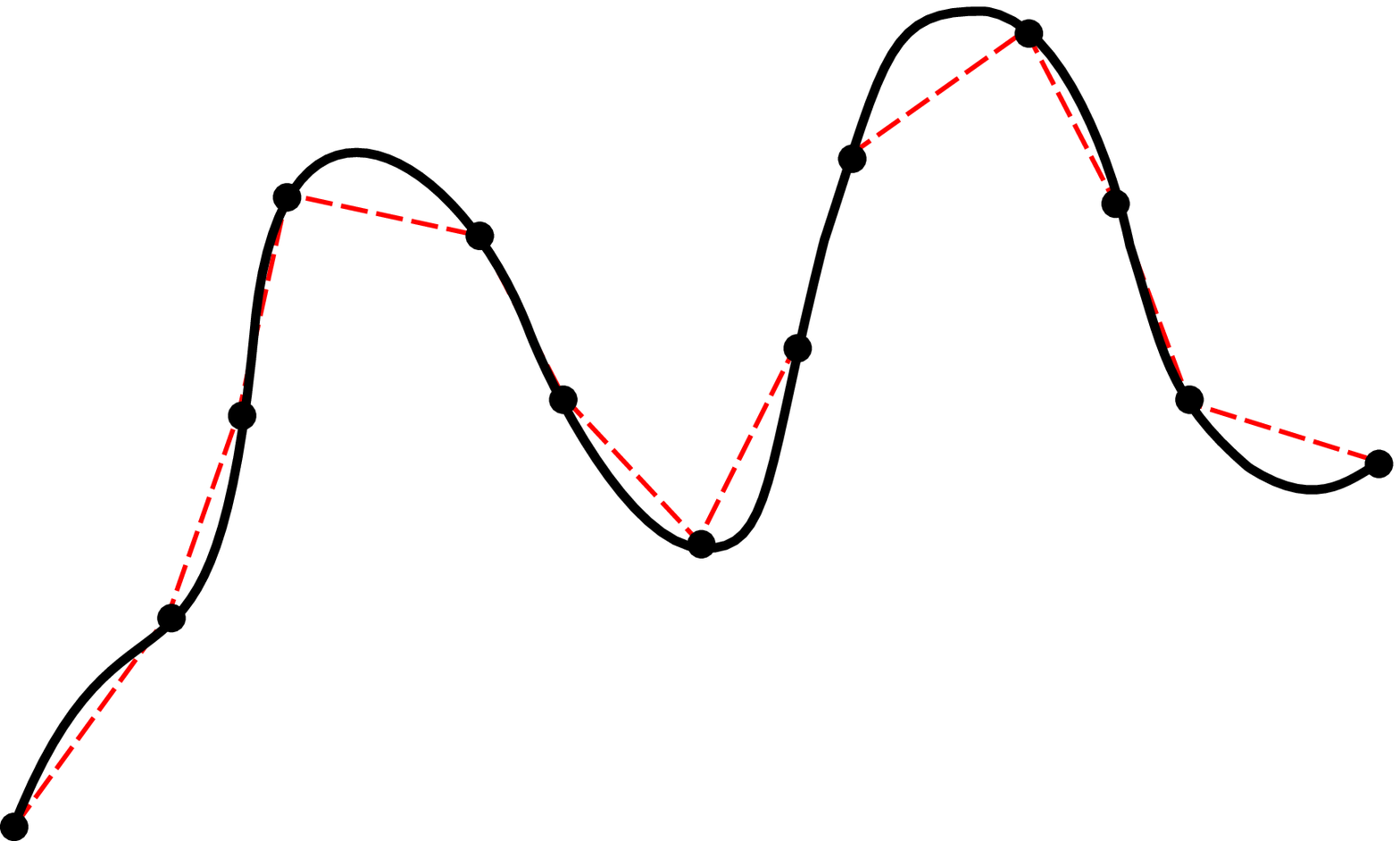}}
\subfloat[]{\includegraphics[height=0.4\textwidth]{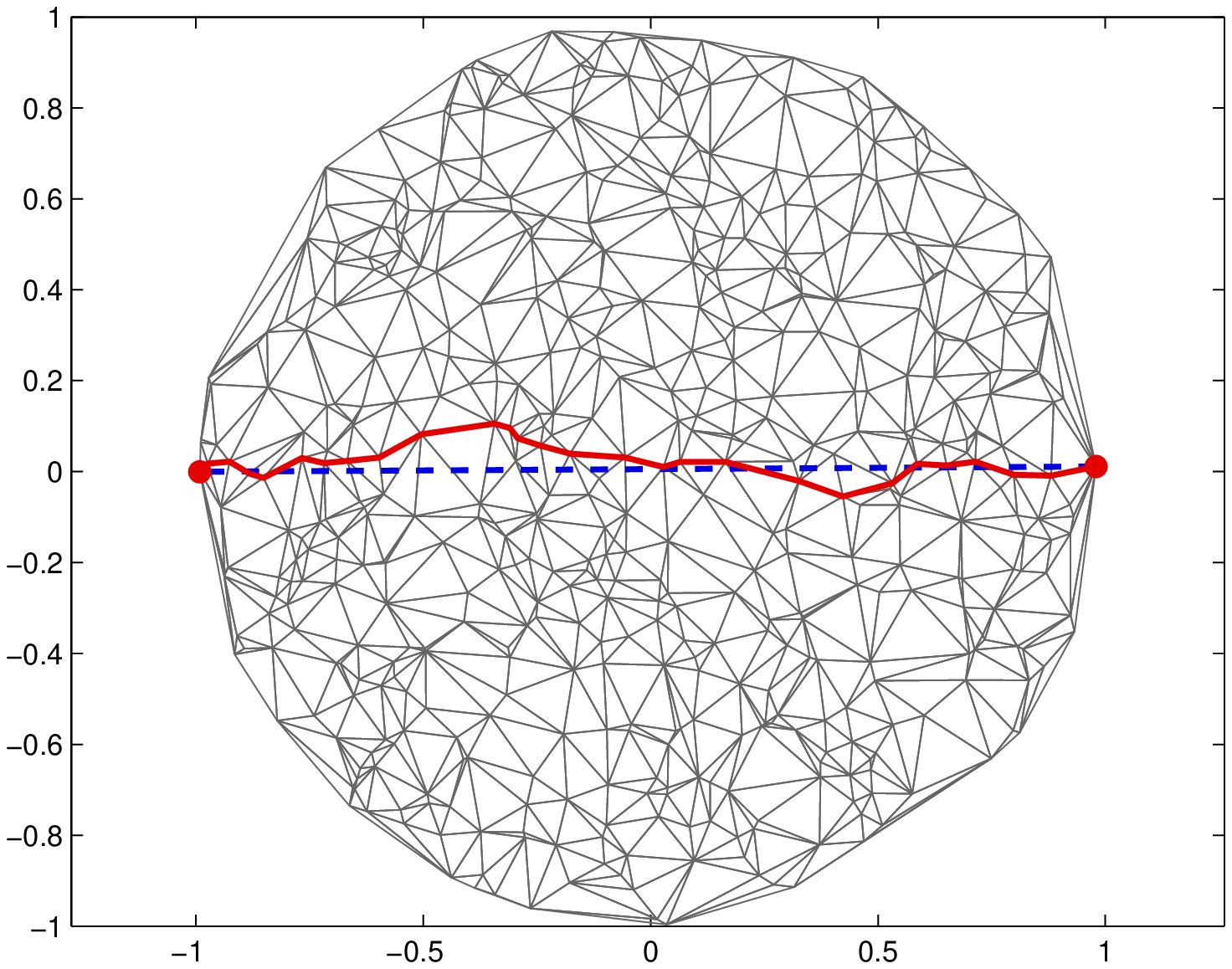}}
\end{center}
\caption{\small (a) Curve representing an arbitrary crack (solid line) and the corresponding rectification (dashed line); (b) shortest euclidean distance and shortest path over the mesh between two random points. }\label{fig:ArbitraryCrack}
\end{figure}

Let us consider a 2-dimensional brittle fracture problem in which an arbitrary crack of length $L_c$ develops in a brittle material, as shown in Figure \ref{fig:ArbitraryCrack}a. The fracture energy $E_f$ required to generate such a crack is proportional to the crack length $L_c$, that is,
\begin{equation}
E_f = 2 \gamma L_c
\end{equation}
where $\gamma$ is the surface energy of the material. This implies that, even assuming a correct model for the surface energy of the material, \emph{the accuracy on any prediction of the fracture energy directly depends on a correct computation of the crack path length}. 

In the most general case, the crack path length could be computed by rectification of the irregular curve, i.e., the crack path could be approximated by connecting a finite number of points on the curve using line segments to create a polygonal path (see Figure \ref{fig:ArbitraryCrack}a). The approximated crack length is then computed as the sum of the length of the corresponding segments. The actual crack length could be then obtained by computing the limit of the approximated length as the segment's length tend to zero. 

When a mesh is introduced to represent the continuum fracture problem within the cohesive element formulation, a constraint is introduced into the problem due to the inability of the mesh to represent, through its edges (2D problems) or faces (3D problems), the shape of an arbitrary crack. By following the reasoning in the preceding paragraph, a necessary condition for the convergence of the cohesive element approach is that the underlying mesh should be able to represent exactly a straight line (2D) or plane (3D) oriented in an arbitrary direction as the mesh is refined \cite{papoulia2006ijnme}. That is, instead of focusing our attention to the problem of representing arbitrary crack shapes, \emph{we could concentrate on the ability of the mesh to represent a straight line (2D) or a plane (3D), and then think of it as a necessary condition to be able to rectificate an arbitrary crack shape}.

In 2-dimensional problems, the ability of a mesh to represent a straight segment is characterized by the path deviation ratio $\eta$, defined as the ratio between the shortest path on the mesh edges connecting two nodes, and the Euclidian distance between them. The path deviation ratio can be interpreted as the maximum relative error in representing a straight line by means of the edges of a finite element mesh. Figure \ref{fig:ArbitraryCrack}b illustrates the relevant quantities to be defined for the computation of the path deviation ratio on a given mesh. This particular mesh was generated by applying Delaunay's triangulation to a set of randomly distributed nodes. The dashed line represents the shortest path over the euclidean manifold between the two points marked in bold. The length of the dashed line is the euclidean distance between those points, $L_e$. On the same figure, the solid line represents the shortest path between the same pair of points over the edges of the mesh as computed by means of Dijkstra's algorithm \cite{dijkstra1959nm}. The sum of the lengths of the segments composing the shortest path defines the length of shortest path on the graph between those points, $L_g$. According to the previous definitions, the path deviation can be computed as
\begin{equation}
\eta=\frac{L_g}{L_e}
\end{equation}

Another associated measure is the relative error in representing a straight line, given by $\epsilon=\eta-1$. In subsequent sections, we may refer to the relative error simply as the error. We may also choose to refer to the behavior of a mesh in terms of the path deviation ratio or in therms of the relative error as we consider necessary for the sake of clarity. Note that decreasing $\eta$ implies decreasing $\epsilon$ and vice versa. Also, note that $\eta \rightarrow 1$ is equivalent to $\epsilon \rightarrow 0$.

Finally, in order to compare path deviation ratios between different kinds of mesh and to compare meshes of different sizes, we introduce the non-dimensional mesh size, defined as
\begin{equation}
\lambda = \frac{\frac{1}{N}\sum\limits_{i=1}^N h_i}{L_e}
\end{equation}
where $N$ is the total number of edges of the given mesh and $h_i$ the length of the $i^{th}$ edge. That is, the non-dimensional mesh size can be interpreted as the inverse of how many \emph{average} segments are needed to represent a segment of length $L_e$. Two meshes with similar values of $\lambda$ would require approximately the same number of edges to represent a given crack.

\subsection{Mesh-induced anisotropy and mesh-induced toughness}

\begin{figure}
\begin{center}
\subfloat[]{\includegraphics[height=0.38\textwidth]{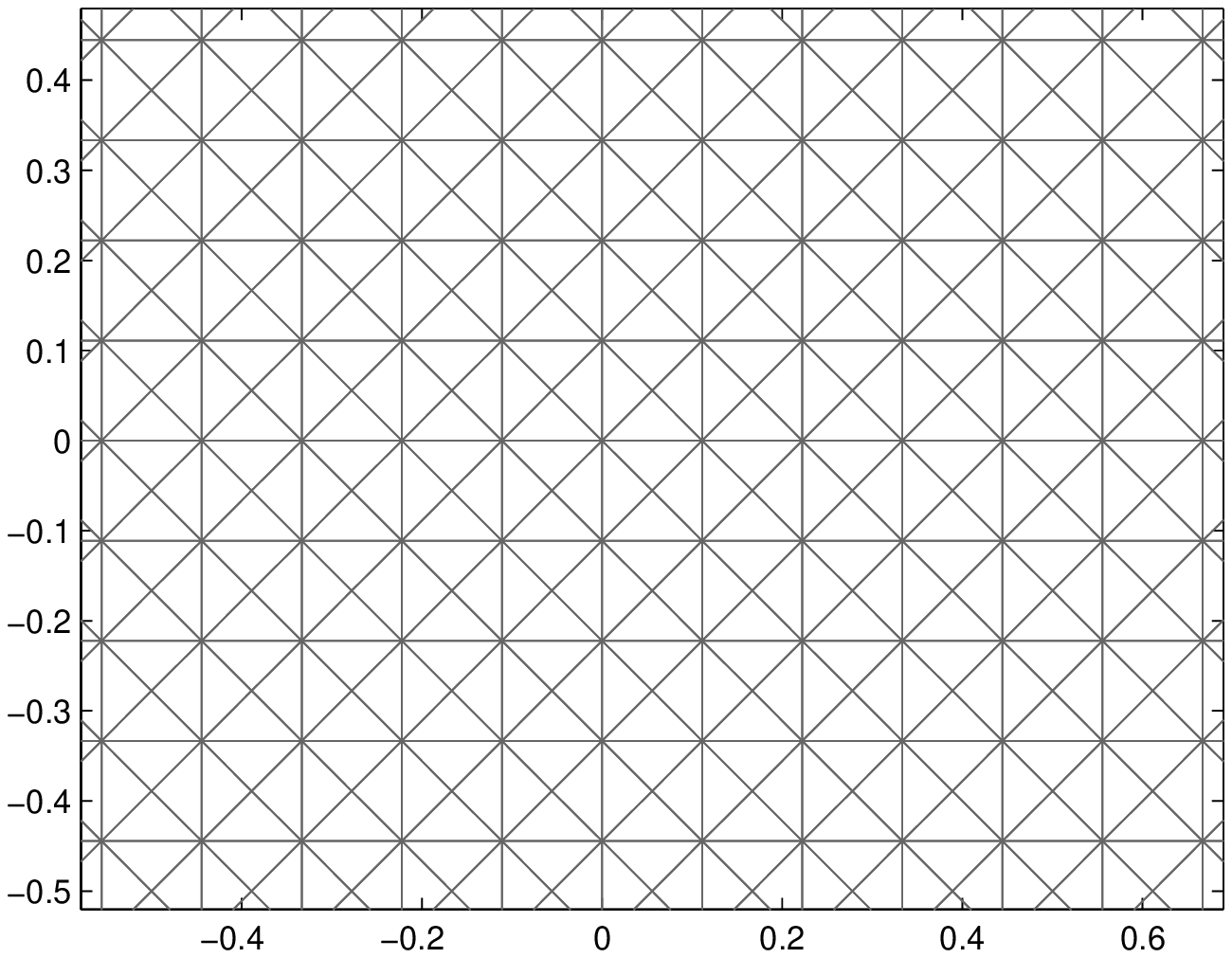}}
\subfloat[]{\includegraphics[height=0.38\textwidth]{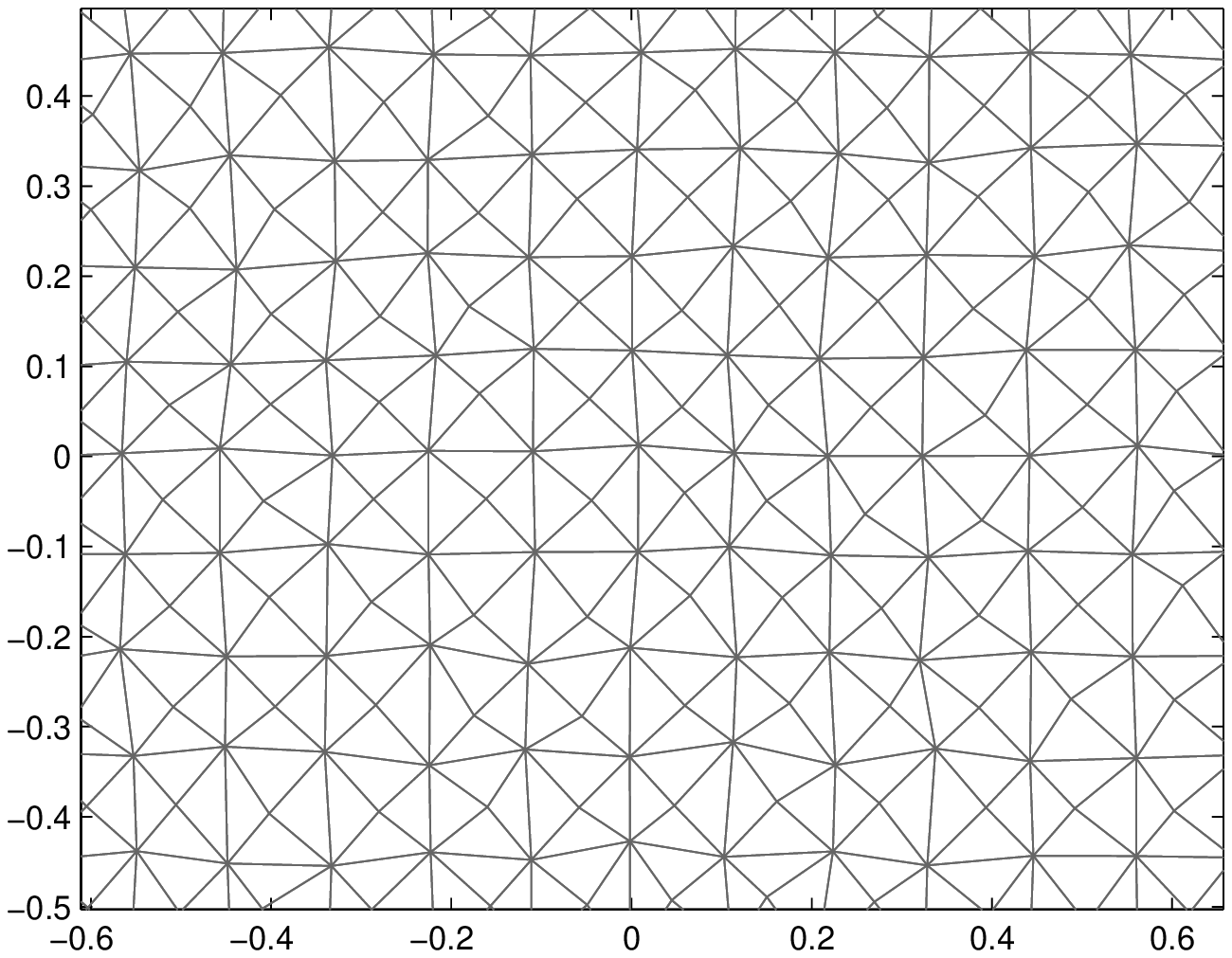}}\\
\end{center}
\caption{\small (a) typical 4k mesh, and (b) 4k mesh with nodal perturbation.}\label{fig:4kMeshes}
\end{figure}

\begin{figure}
\begin{center}
\subfloat[]{\includegraphics[height=0.48\textwidth]{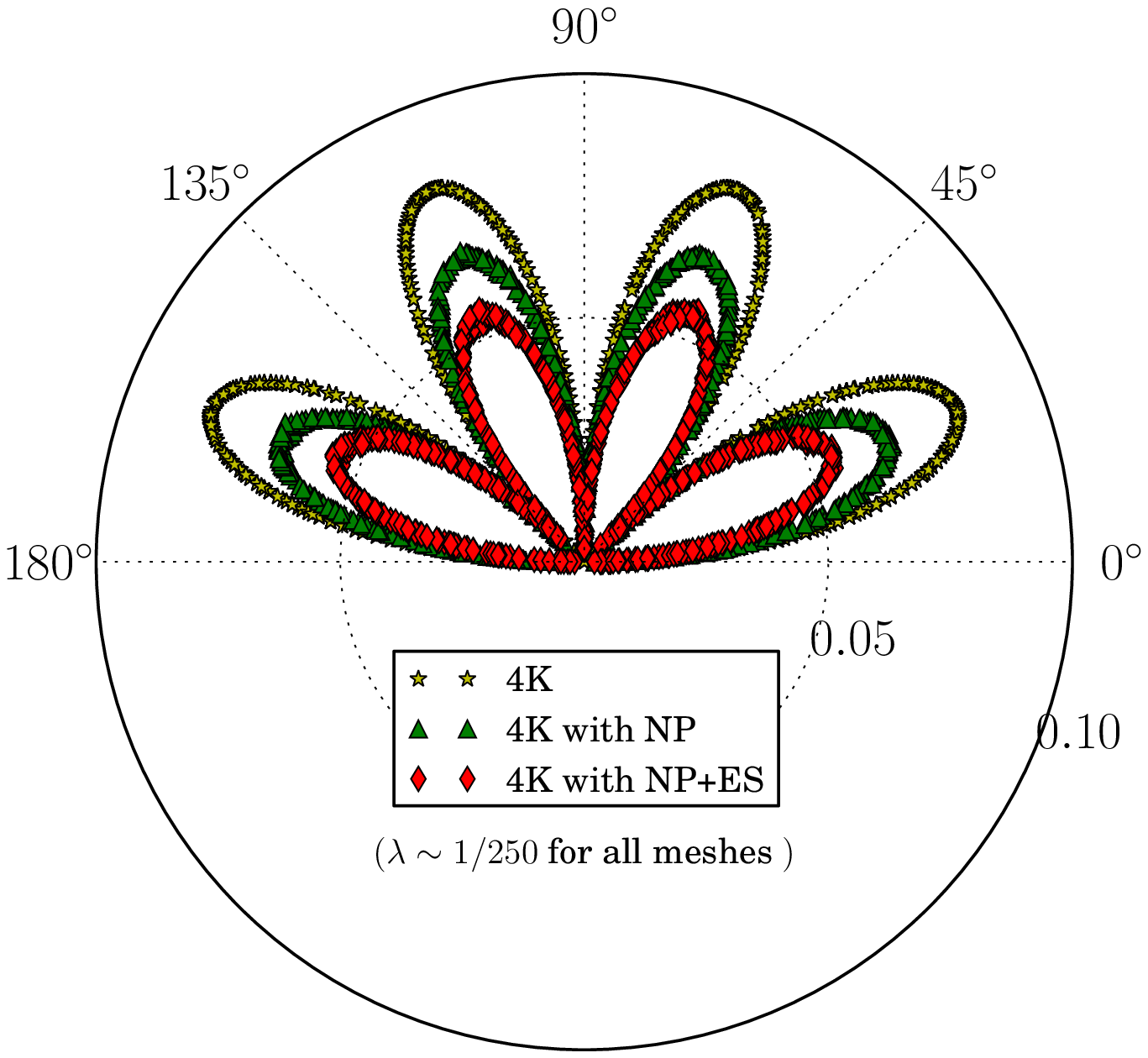}}
\subfloat[]{\includegraphics[height=0.48\textwidth]{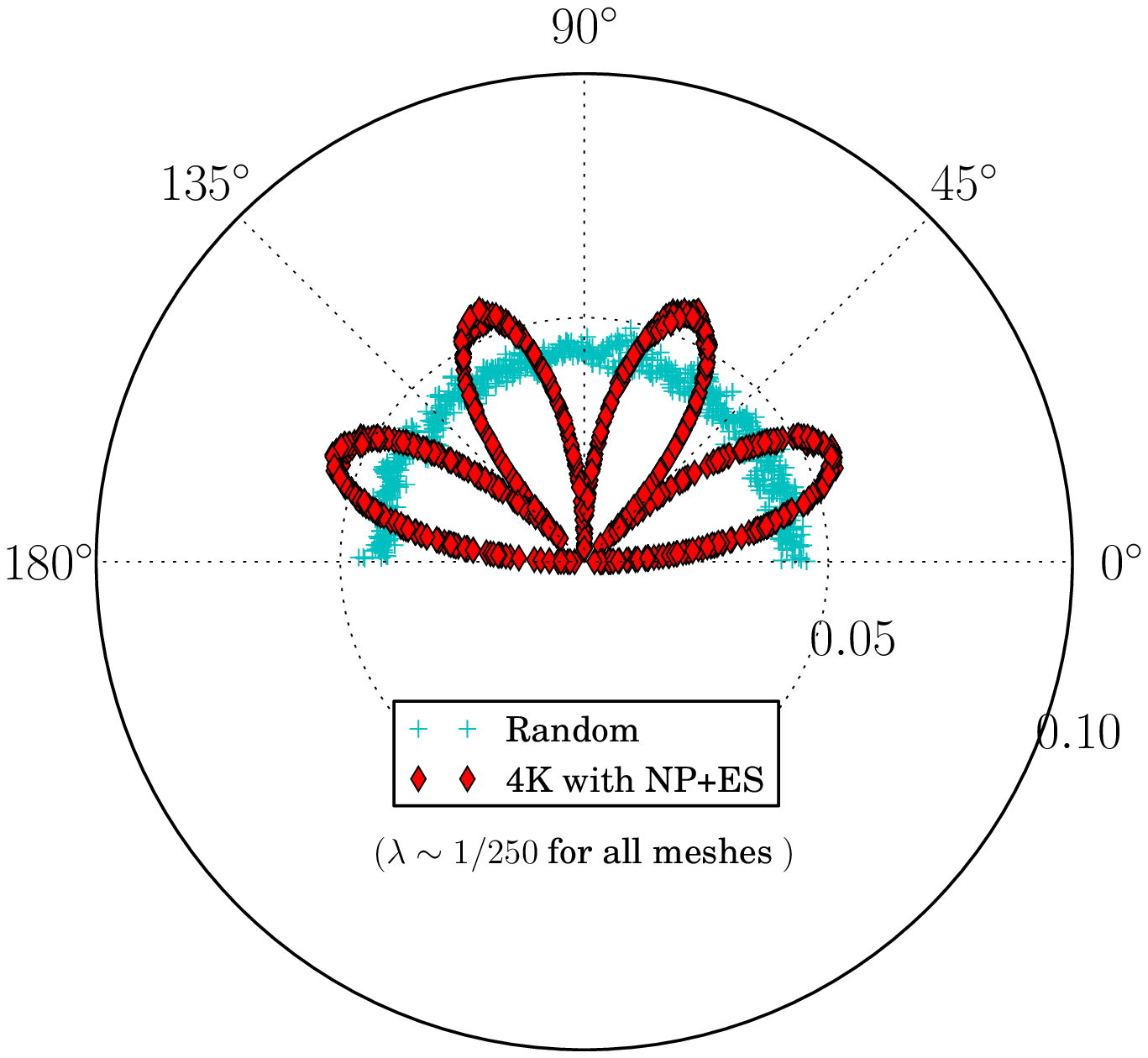}}
\end{center}
\caption{\small Relative error ($\epsilon=\eta-1$) vs mesh direction for (a) 4k mesh, 4k mesh with nodal perturbation, and 4k mesh with nodal perturbation and edge swap, and (b) 4k mesh with nodal perturbation and edge swap, and random mesh.}\label{fig:4kError}
\end{figure}

By definition, the path deviation ratio is greater or equal than one, i.e., $\eta \ge 1$. Consequently, an arbitrary crack can only be approximated by a discrete curve of equal or larger length. For purposes of illustration, let us consider the 4k mesh depicted in Figure \ref{fig:4kMeshes}a. The diamond symbols on Figure \ref{fig:4kError}a show, for such a mesh with $\lambda\approx 1/200$, how the relative error $\epsilon$ changes as function of a chosen direction on the mesh. As observed in the figure, the error $\epsilon=0$ for the horizontal, vertical, and $\pm45^{\circ}$ directions. However, it reaches a peak value near 0.08 for intermediate directions. In energetic terms, directions for which the relative error (or path deviation ratio) is lower provide favorable paths for crack propagation. Clearly, cracks will tend to align along those lower-energy directions leading to inaccurate numeric predictions. We term this effect \emph{mesh-induced anisotropy}.

In general, when looking at all possible directions at a given point on the mesh, the mean value of the path deviation ratio is strictly larger than one, i.e.,  $\text{\bf mean}(\eta) >1 $. This, in turn, implies that the total energy released by the discrete representation of an arbitrary crack $E_{fg}$ is larger than the  energy $E_f$ released by a real crack on the continuum. That is, the introduction of a discrete mesh necessarily leads to a larger energy dissipation by the discrete cohesive model. We call this effect \emph{mesh-induced toughness}.

A necessary condition to avoid the undesired mesh-induced anisotropy is for the path deviation ratio $\eta$ (or relative error $\epsilon$) to be independent on the chosen direction. Additionally, a necessary condition to avoid the undesired mesh-induced toughness, is for the path deviation ratio to tend to 1 (or equivalently for the relative error to tend to 0) for all possible directions in the mesh as the non-dimensional mesh size $\lambda$ tends to zero. The satisfaction of these two conditions is usually referred to as isoperimetric property. The work of Radin and Sadun \cite{radin1996cmp} shows that the pinwheel tiling of the plane has the isoperimetric property. Based on this result, Papoulia et al. \cite{papoulia2006ijnme} observed that crack paths obtained from pinwheel meshes are more stable as meshes are refined when compared to other types of meshes. It is worth noting, however, that pinwheel meshes are difficult to generate and that there is no known extension of the pinwheel tiling to the 3-dimensional case.

Recently, Paulino et al. \cite{paulino2010ijnme} introduced two mesh operators that decrease the undesired mesh dependent effects observed in 4k meshes. First, they introduced a nodal perturbation operator (NP operator) that randomly moves the nodes of 4k meshes by an amount proportional to the length of the shortest edge in the corresponding 4k cell. For example, Figure \ref{fig:4kMeshes}b shows a typical 4k mesh affected by the nodal perturbation operator. Subsequently, they introduced a topological edge swap operator (ES operator) that allows the originally vertical edges of the 4k mesh to switch to a horizontal position and vice versa. The idea behind these operators is to break the symmetry of the mesh at the geometrical level (NP operator) and to relax its structure at the topological level (ES operator).

As noted in their work, both operators produce an increment in the value of the path deviation ratio (or relative error) in the directions that showed no error for the 4k mesh, while at the same time they reduce it for the intermediate directions. They also show that the reduction of the path deviation ratio in the intermediate directions is larger than the increment produced in the horizontal, vertical, and $\pm45^{\circ}$ directions. The combination of these effects results in an effective reduction of the mean value of the relative error $\text{\bf mean}(\epsilon)$. It is worth noting, however, that even though the mesh induced-toughness is reduced in this way, the meshes under consideration still exhibit high anisotropy, \emph{only evident when considering the full polar plot of the relative error (or path deviation ratio)}. We illustrate this effect by analyzing a 4k mesh with nodal perturbation factor $\text{NP}_f=0.3$ (4k with NP in Figure \ref{fig:4kError}a), and a 4k mesh perturbed by the same amount with the extra possibility of swapping edges (4k with NP+ES in Figure \ref{fig:4kError}a). As Figure \ref{fig:4kError}a shows, for meshes with non-dimensional mesh size $\lambda \approx 1/250$,  the maximum attained value of the path deviation ratio is decreased by the application of the NP operator. Furthermore, further application of the ES operator results in even lower values of the path deviation ratio. However, for all three cases the path deviation ratio highly depends on the direction under consideration. That is, both the NP and ES operators seem to have little effect on the mesh-induced anisotropy.

To conclude this section, we observe an interesting feature that arises from analyzing the polar behavior of a random mesh, i.e., a mesh generated through the application of Delaunay's triangulation algorithm to a set of randomly placed points. Figure \ref{fig:4kError}b depicts the relative error dependence on the direction for a random mesh and a 4K mesh subject to the NP and ES operators. The adopted value of the nodal perturbation factor is $\text{NP}_f=0.3$, and both meshes have comparable values of $\lambda\approx 1/250$. It is interesting to note that whereas the 4k mesh exhibits a lower mean value of the relative error ($\text{\bf mean}(\epsilon)\approx 0.037$ vs $\text{\bf mean}(\epsilon)\approx 0.043$), the random mesh shows a much better isotropic behavior. Needles to say, we would ideally like to have a mesh that reflects a compromise between these two behaviors, that is, a low relative error that remains constant for all possible directions.

\section{K-means and conjugate-directions meshes}\label{sec:CDM}

\subsection{Isotropy through K-means meshes}
Based on the previous observations, we wish to generate random meshes in order to achieve isotropy in the sense of the path deviation ratio. At first, purely random meshes look attractive: even though they generate path deviation ratios with mean value slightly higher than those obtained by 4k meshes under the effects of NP and ES operators, they have the advantage of not exhibiting any undesired anisotropy. However, purely random meshes are very irregular and present highly distorted triangles rendering them unfit for finite element computations. To address this issue, we propose to \emph{smooth} purely random meshes by subjecting them to a K-means clustering algorithm. The K-means clustering algorithm partitions $spn\cdot n$ observations into $n$ clusters in which each observation belongs to the cluster with the nearest mean. Then, a so called K-means mesh can be obtained as follows:
\begin{itemize}
\item $spn\cdot n$ uniformly random points are clustered into $n$ nodes.
\item The resulting nodes are then triangulated by means of Delaunay's algorithm.
\end{itemize}

\begin{figure}
\begin{center}
\includegraphics[height=0.4\textwidth]{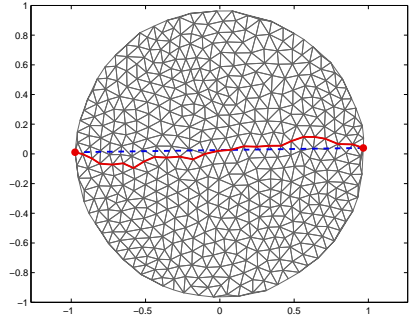}
\end{center}
\caption{\small Typical K-means mesh obtained by clustering $128\cdot 512=65536$ random points into $512$ nodes ($spn=128, n=512$).}\label{fig:Kmeans}
\end{figure}

Details of the k-means clustering algorithm used in this article are depicted in Algorithm \ref{alg:KMeans} for completeness. For more details on the k-means algorithm and its relationship to centroidal Voronoi tesselations, see \cite{du1999sr}. Figure \ref{fig:Kmeans} shows a typical K-means mesh obtained obtained by clustering $128\cdot 512=65536$ random points into $512$ nodes ($spn=128, n=512$).

\begin{algorithm}
\KwIn{List $\{\mathbf z_i\}_{i=1}^n$ of $n$ uniformly distributed random points (initial cluster positions).}
\KwIn{List $\{w_i\}_{i=1}^n$ of $n$ integers with $w_i=1$ for all i (initial cluster weights).}
\KwIn{List $\{\mathbf y_j\}_{j=1}^{n \cdot (spn-1)}$ of $(spn-1)\cdot n$ uniformly distributed random points.}
\KwOut{List $\{\mathbf z_i\}_{i=1}^n$ of $n$ final cluster positions.}
\tcc{Iterate over clustering points}
\For{each $\mathbf y_j$}
{
	find the $\mathbf z_i$ that is closest to $\mathbf y_j$ and denote the index of that $\mathbf z_i$ by $i^*$;\\
	$\mathbf z_{i^*} \leftarrow \frac{w_{i^*} \mathbf z_{i^*} + \mathbf y_{j^*}}{w_{i^*}+1}$;\\
	$w_{i^*} \leftarrow w_{i^*} + 1$;
}

\KwRet{$\{\mathbf z_i\}_{i=1}^n$}
\caption{K-means clustering.}
\label{alg:KMeans}
\end{algorithm}

To better illustrate the effect of the K-means clustering algorithm on random meshes, we demonstrate the influence of the clustering parameter $spn$ on both the spatial node distribution, and on the triangle quality distribution of the resulting meshes.

\begin{figure}
\begin{center}
{\small
$\text{spn}=1$ \\(uniformly random distribution)}\\
\subfloat{\includegraphics[height=0.28\textwidth]{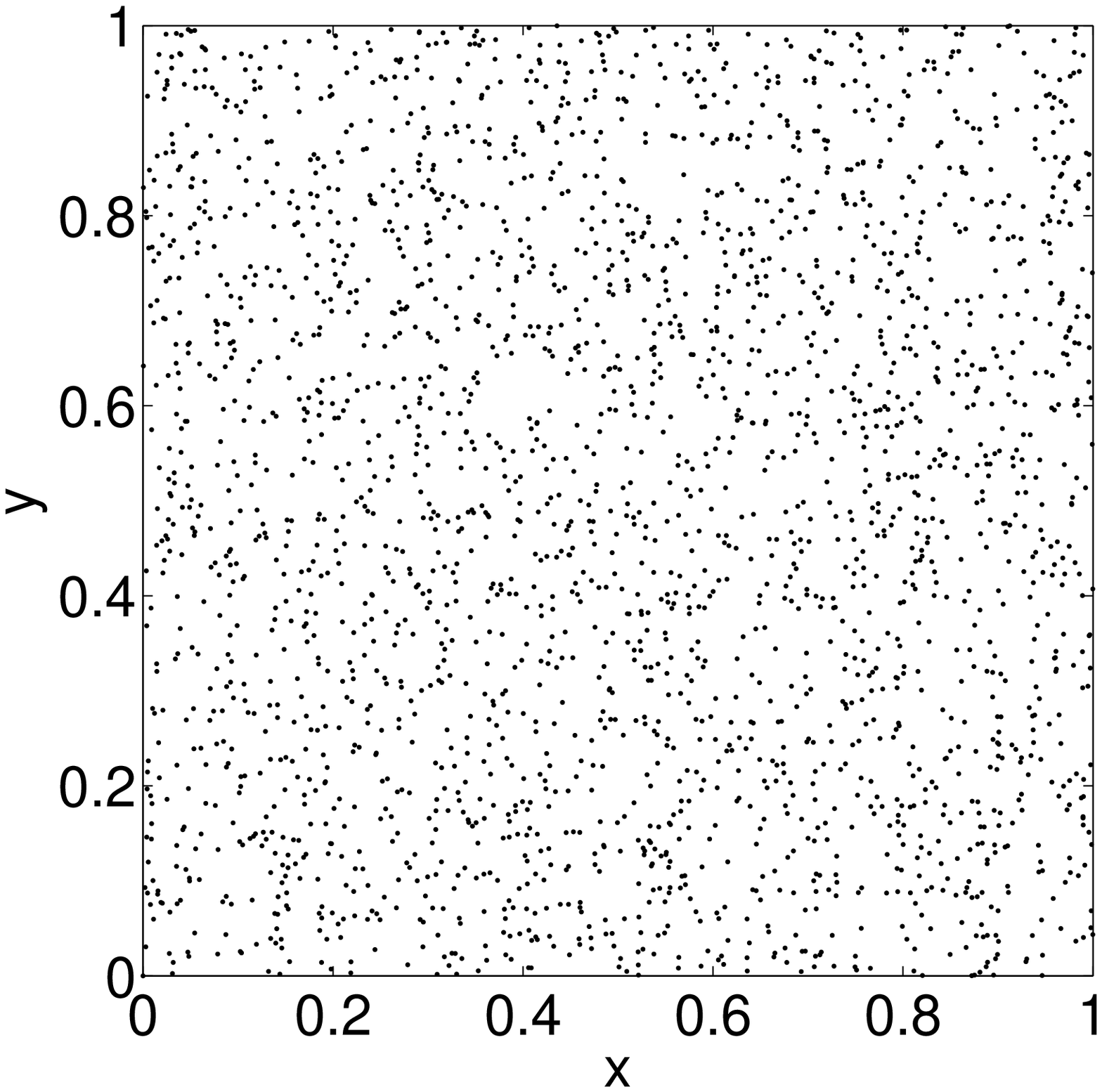}}
\subfloat{\includegraphics[height=0.28\textwidth]{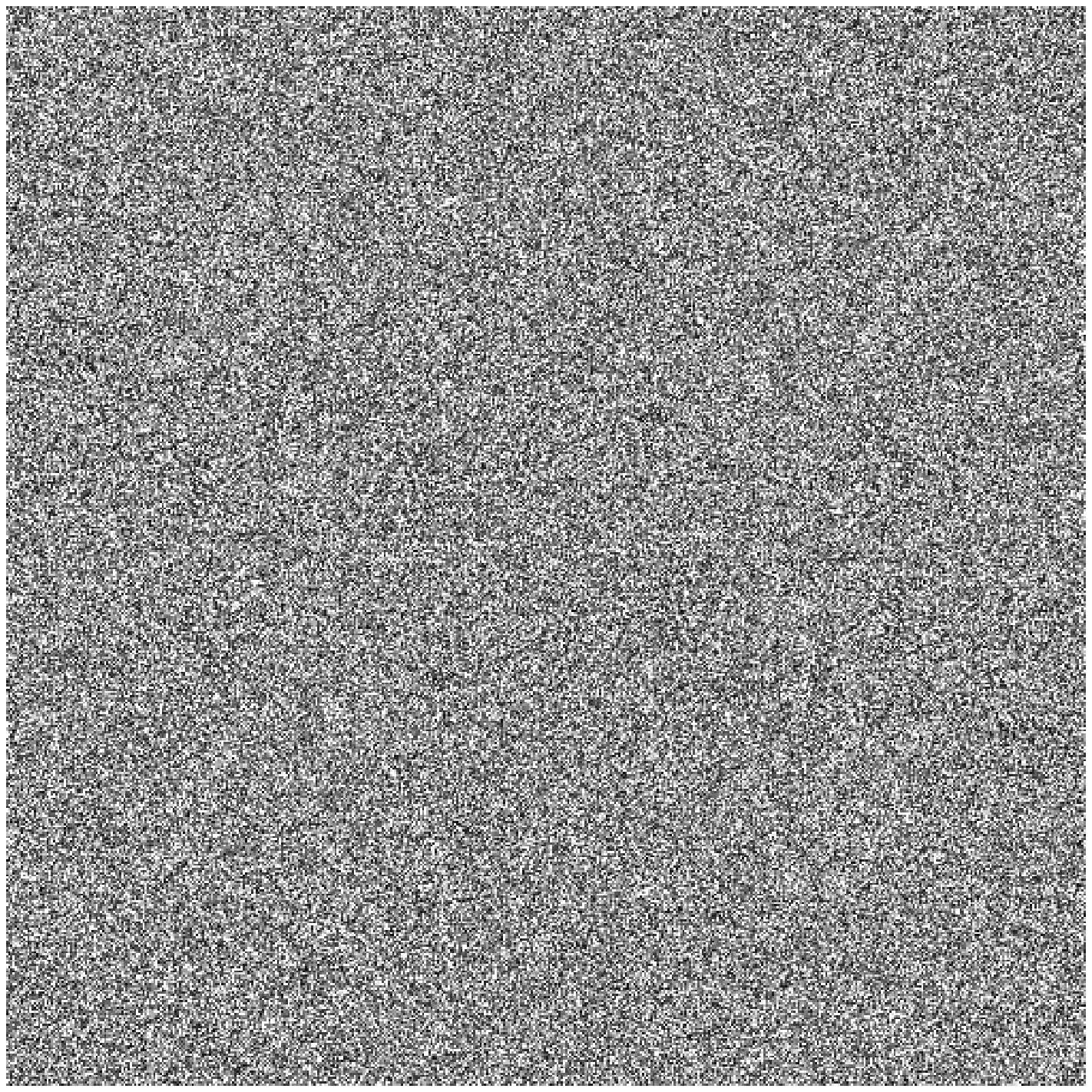}}
\subfloat{\includegraphics[height=0.28\textwidth]{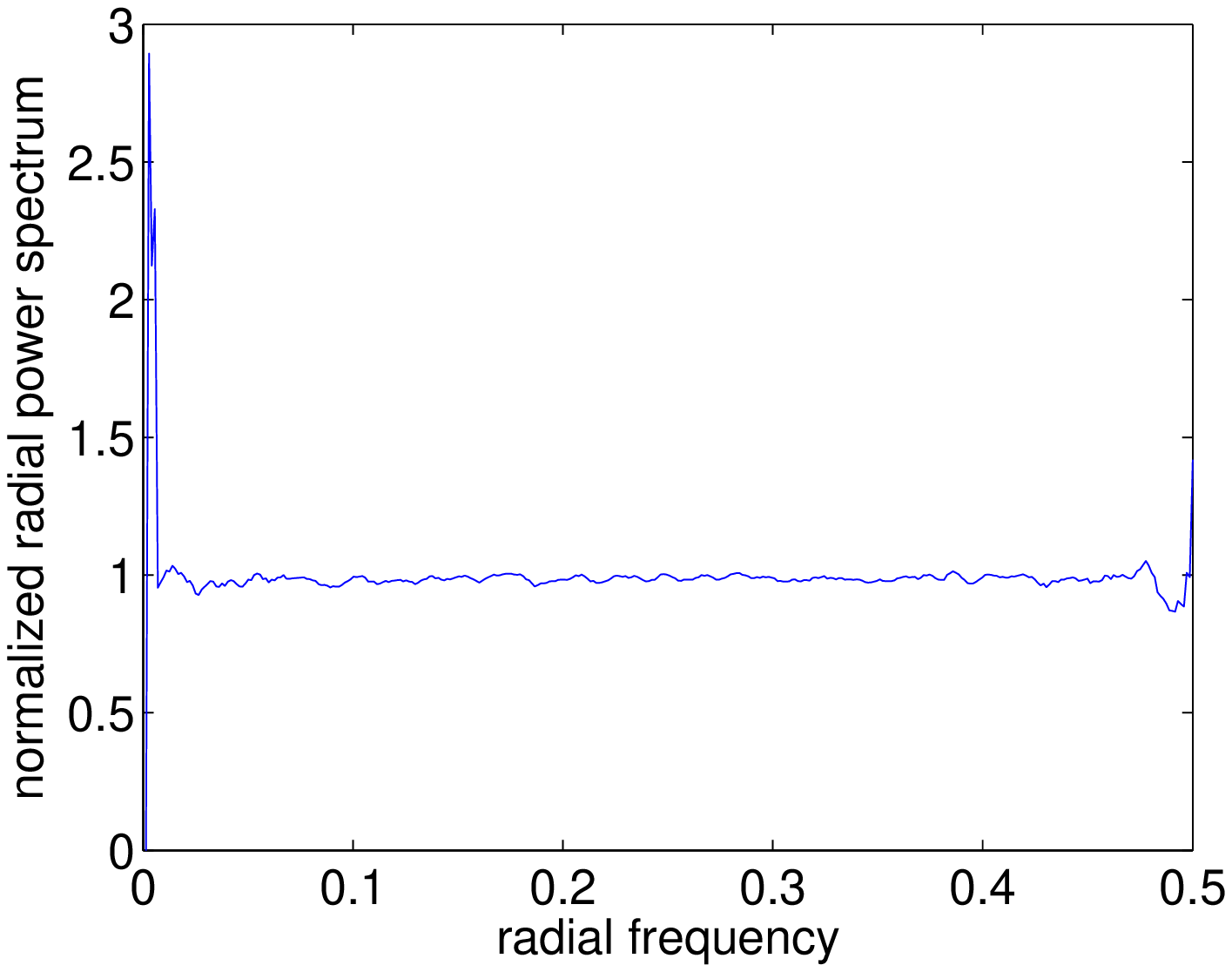}}\\
{\small
$\text{spn}=8$} \\
\subfloat{\includegraphics[height=0.28\textwidth]{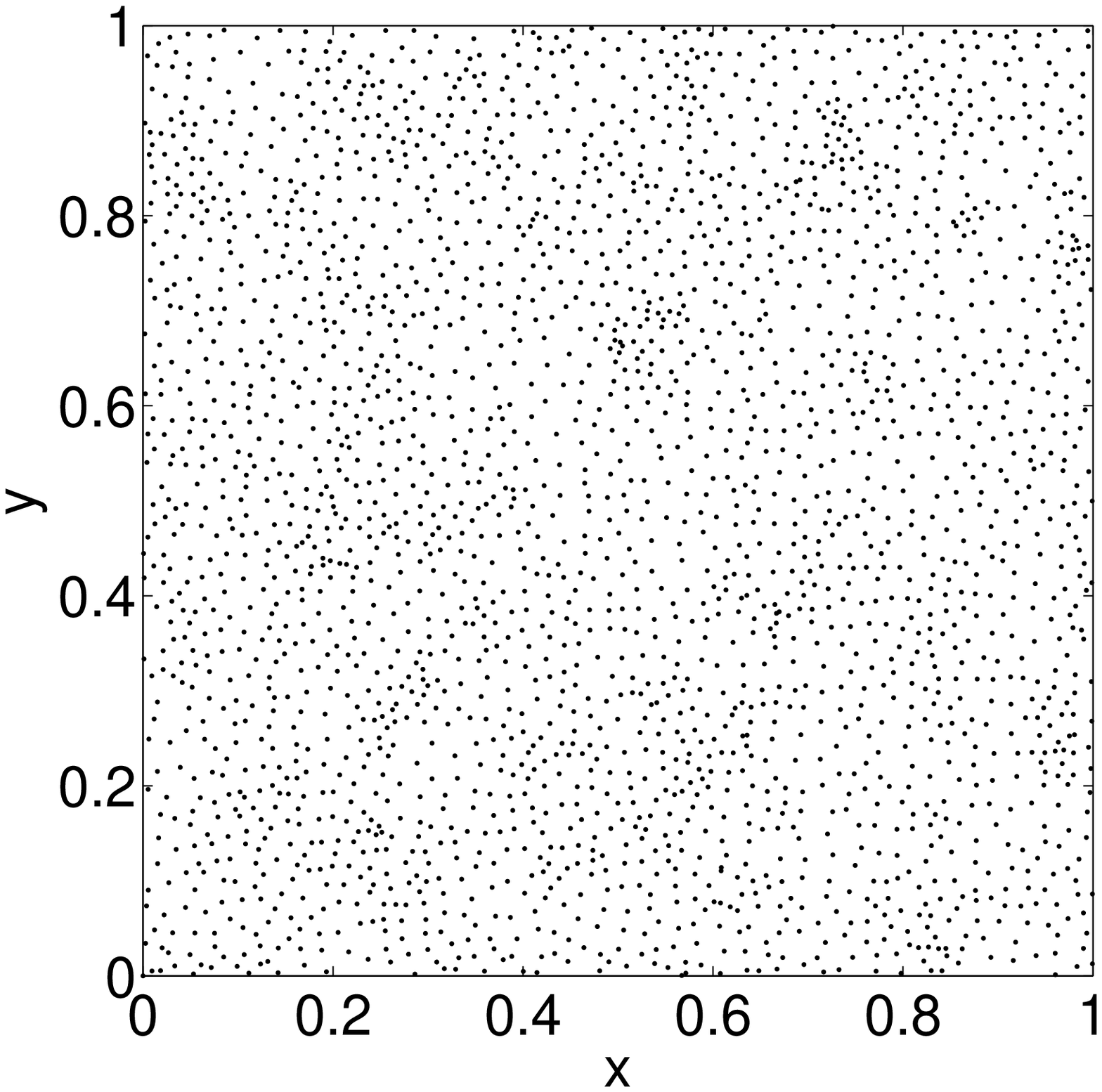}}
\subfloat{\includegraphics[height=0.28\textwidth]{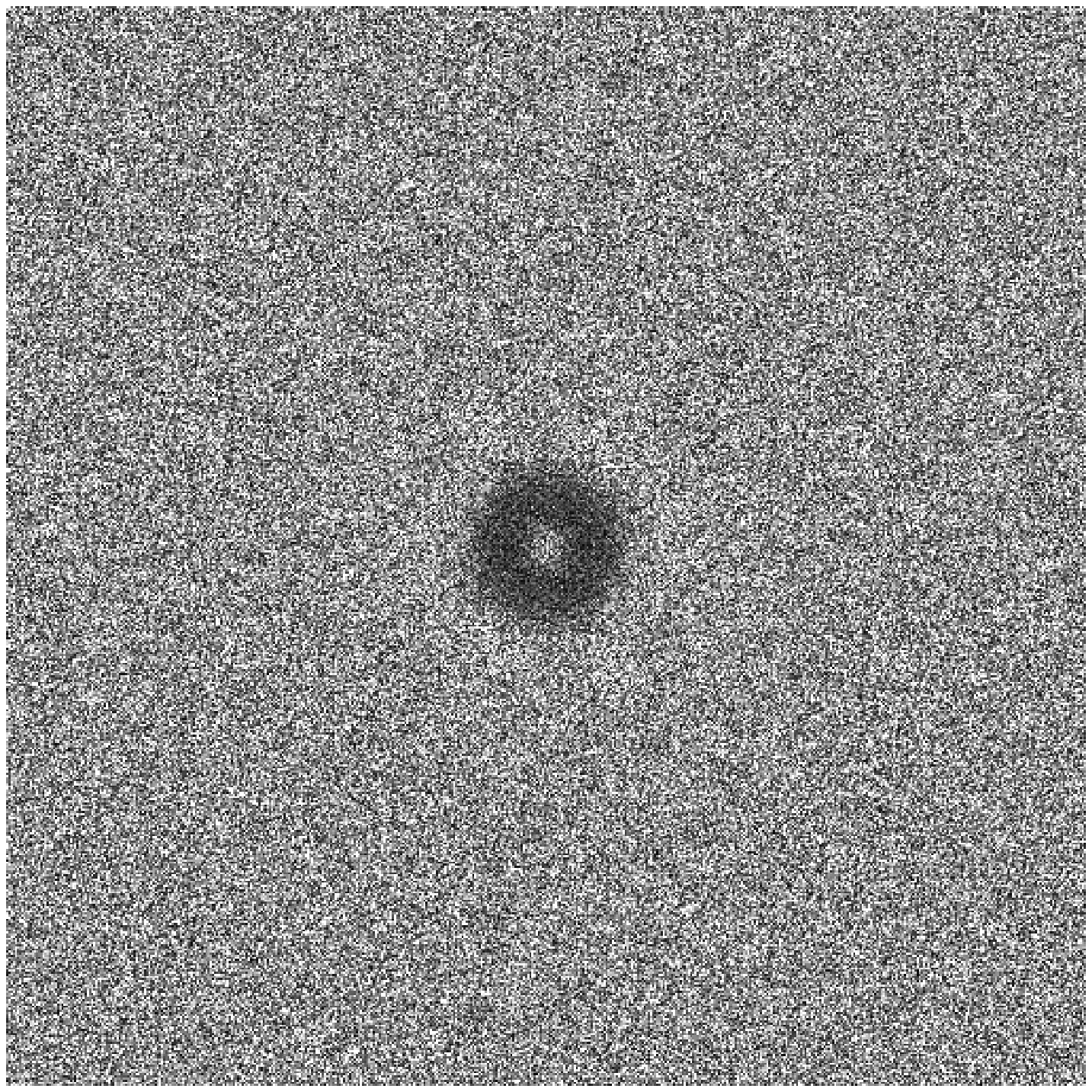}}
\subfloat{\includegraphics[height=0.28\textwidth]{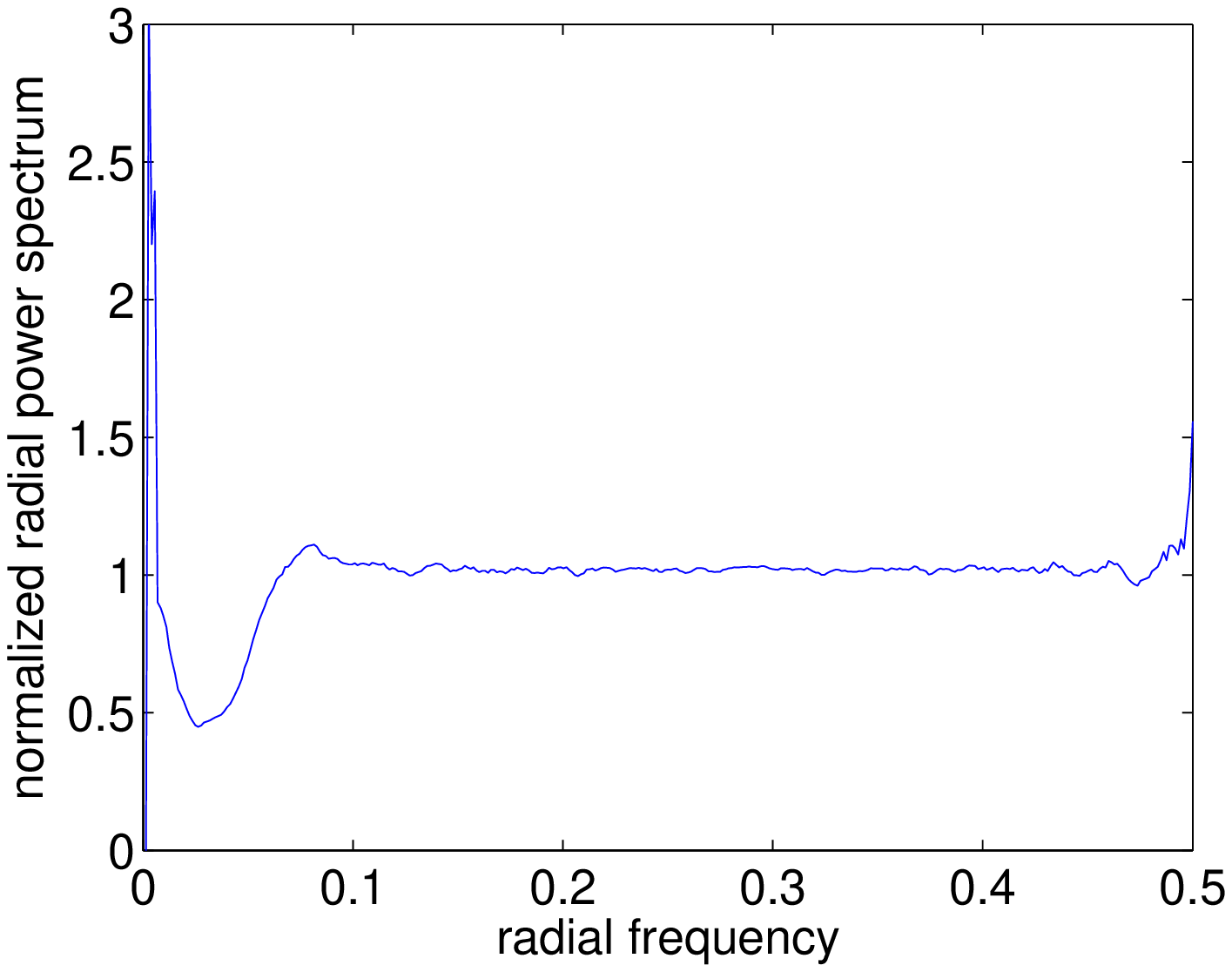}}\\
{\small
$\text{spn}=32$} \\
\subfloat{\includegraphics[height=0.28\textwidth]{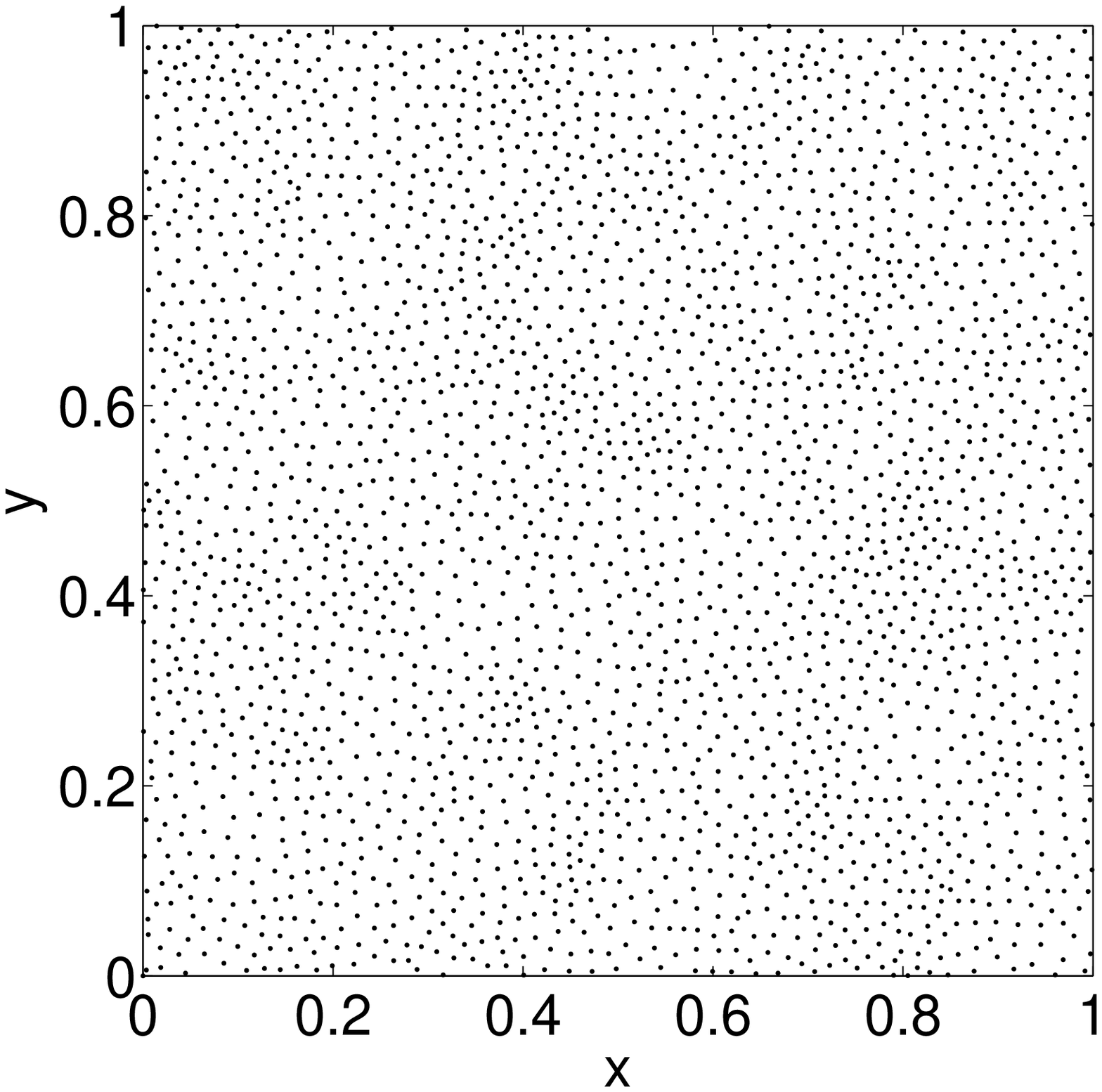}}
\subfloat{\includegraphics[height=0.28\textwidth]{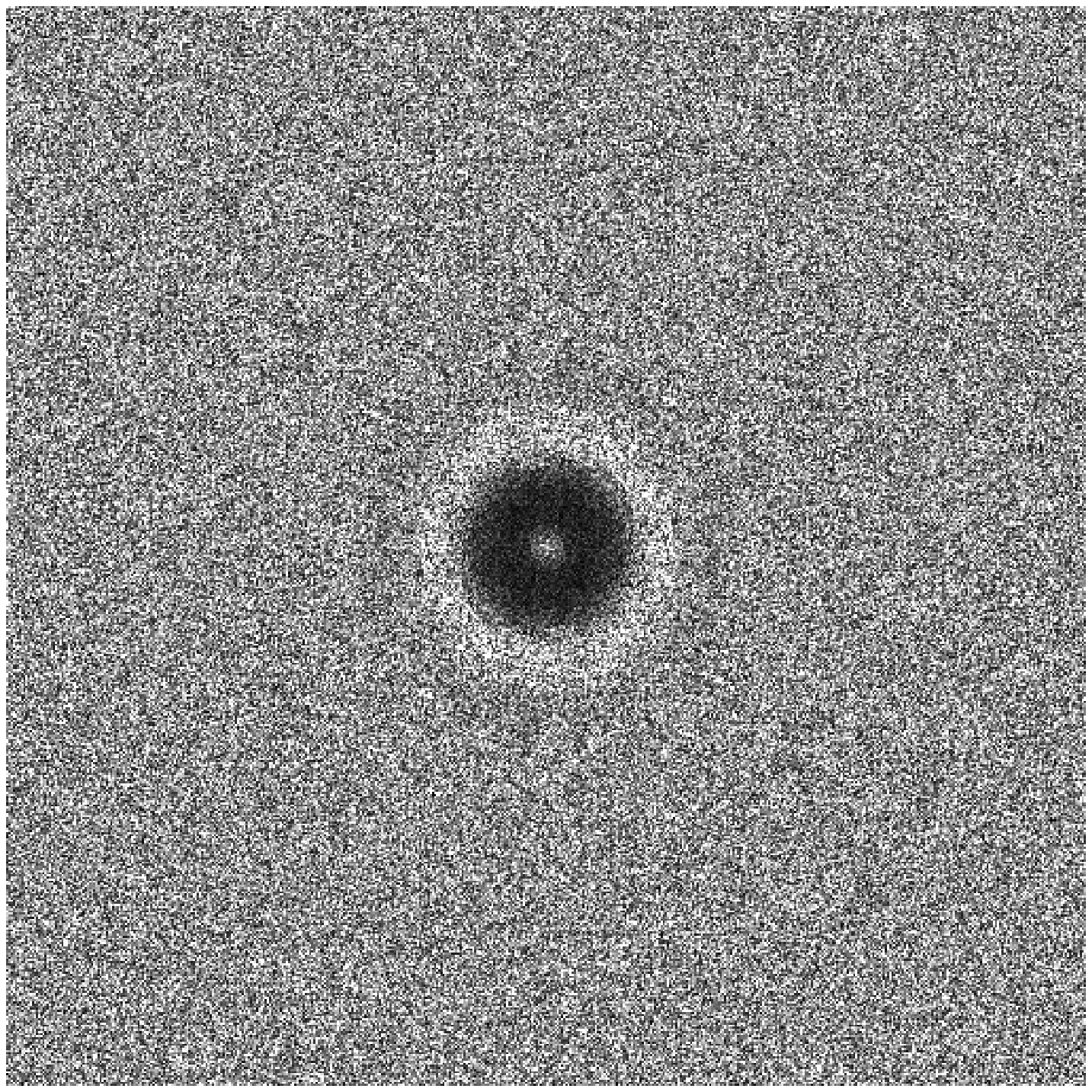}}
\subfloat{\includegraphics[height=0.28\textwidth]{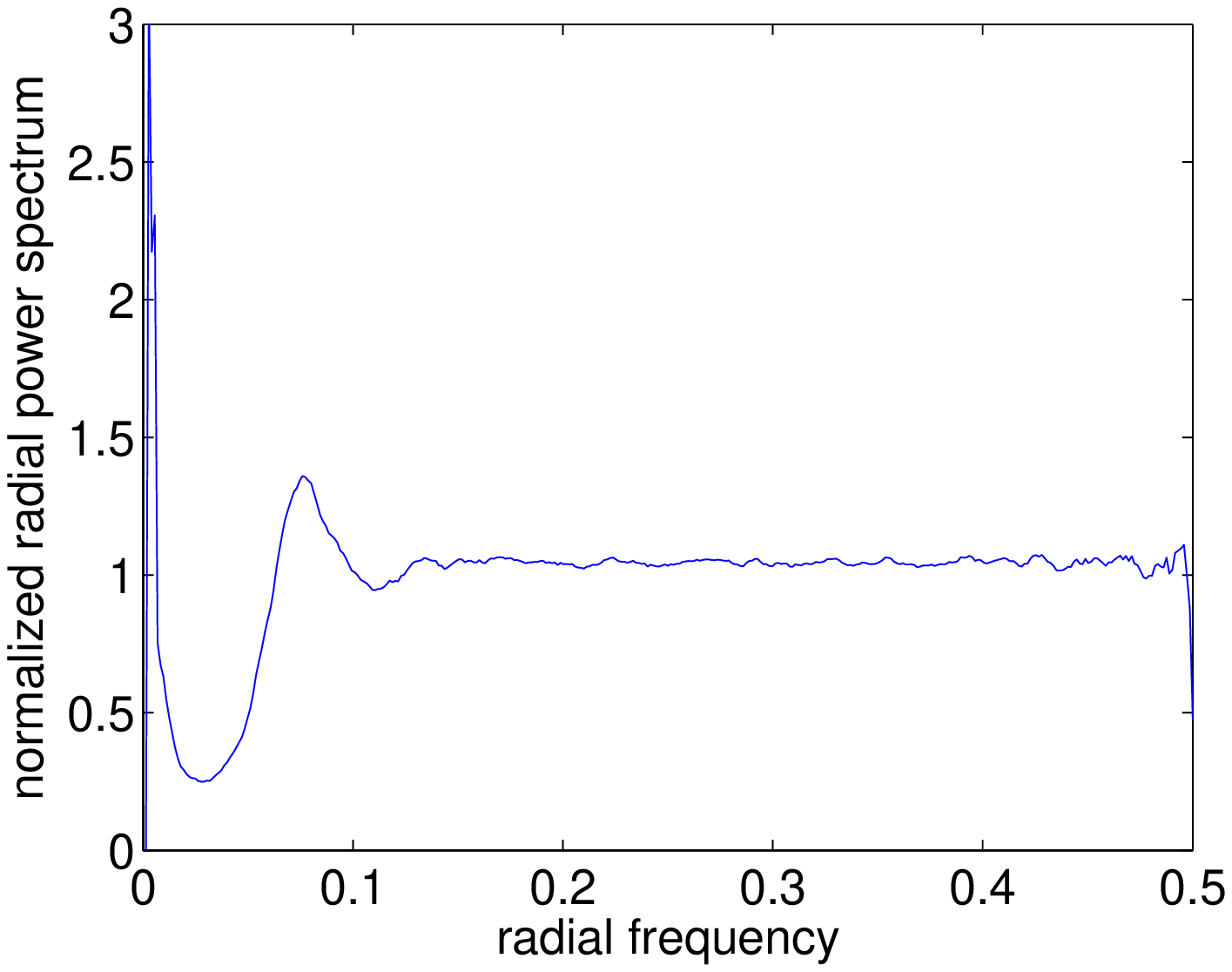}}\\
{\small
$\text{spn}=128$} \\
\subfloat{\includegraphics[height=0.28\textwidth]{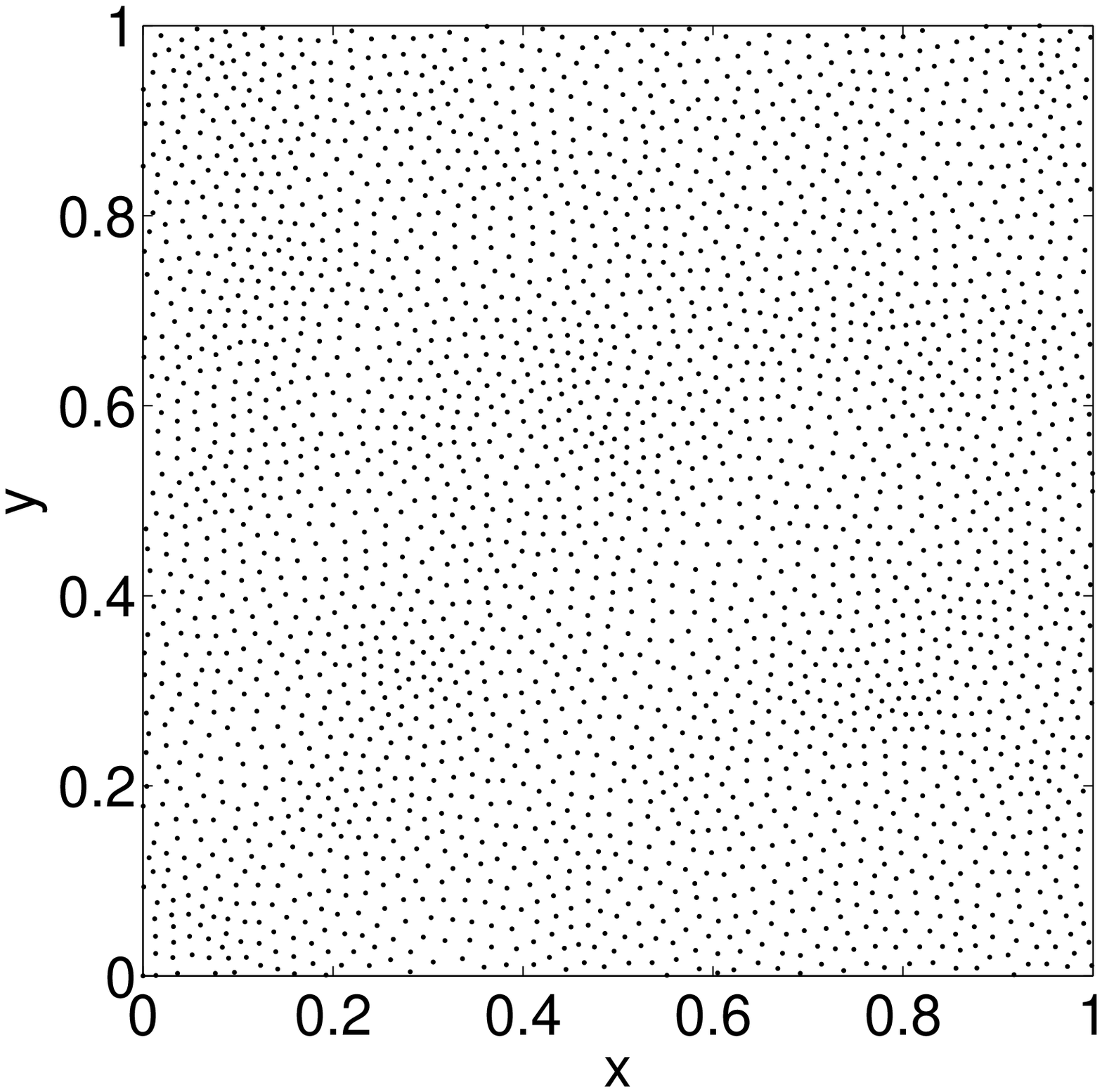}}
\subfloat{\includegraphics[height=0.28\textwidth]{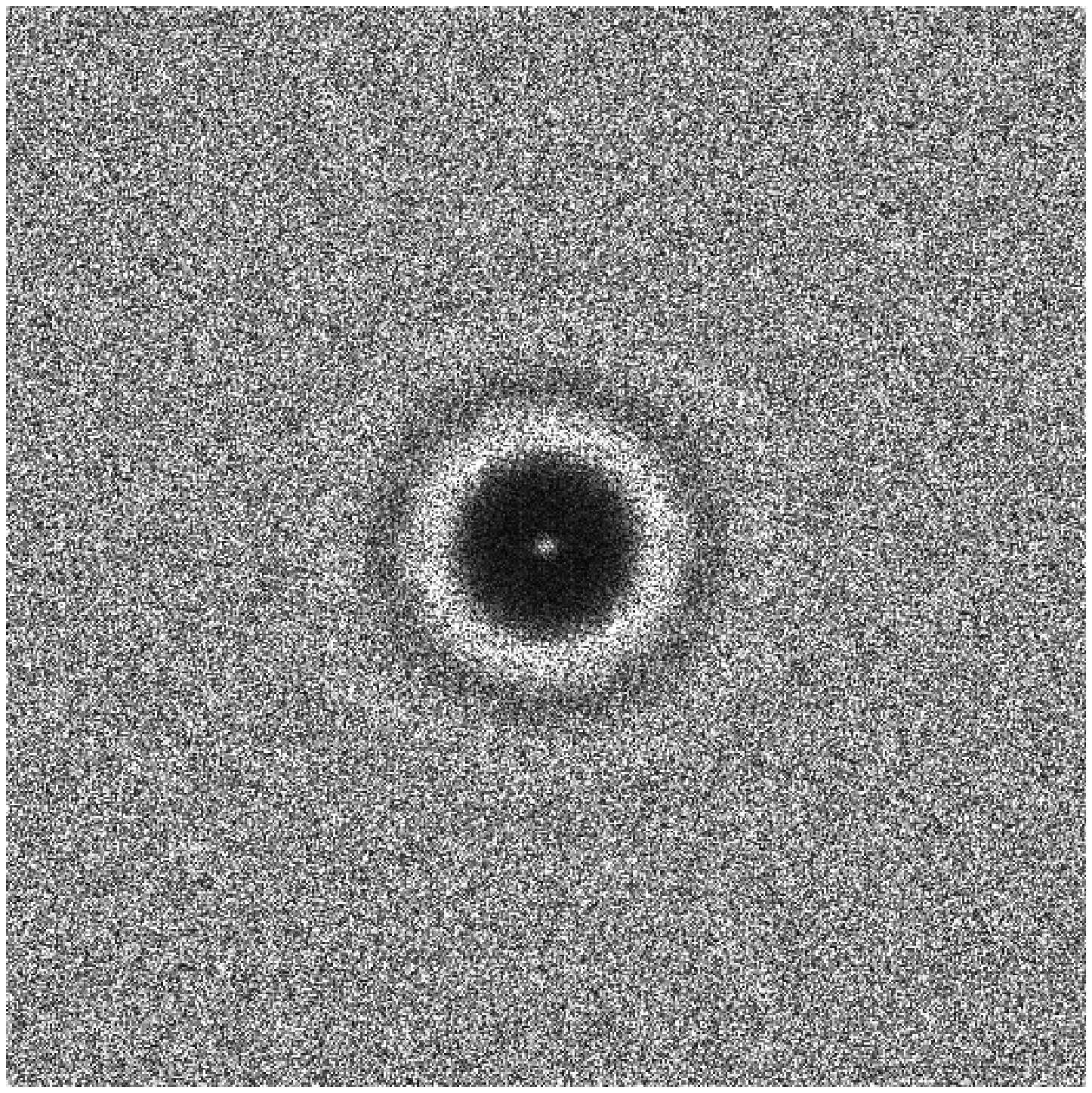}}
\subfloat{\includegraphics[height=0.28\textwidth]{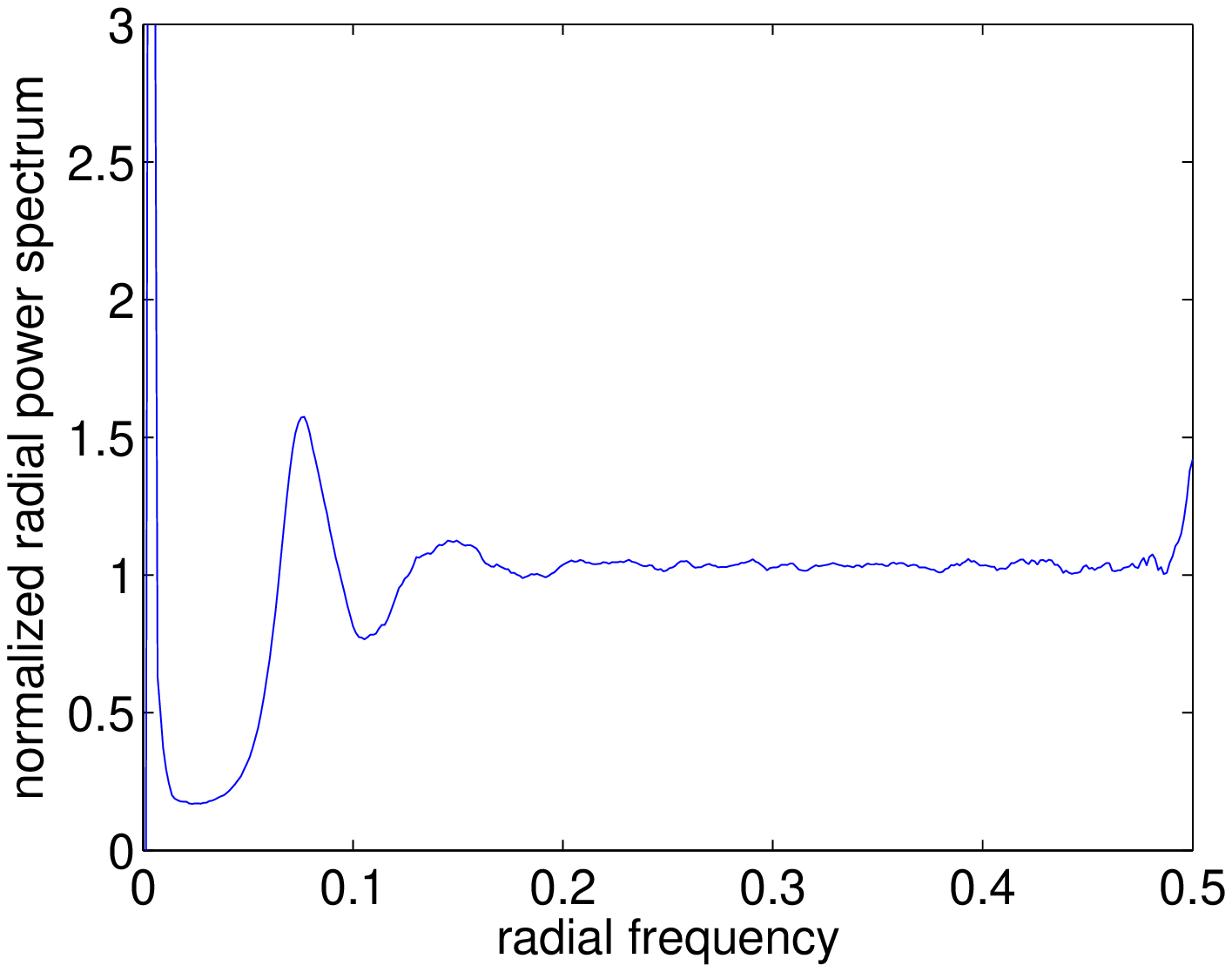}}\\
\end{center}
\caption{\small Influence of the smoothing parameter $spn$ on mesh node distribution. Spatial node distribution (left column), 2D-Fourier power spectrum (center column), and radially averaged power spectrum (right column).}\label{fig:KMeansSpectrum}
\end{figure}

Figure \ref{fig:KMeansSpectrum} shows the effect of the clustering parameter $spn$ on the spatial distribution of mesh nodes and the corresponding Fourier spectrum. The Figure is organized in a matrix layout where rows (from top to bottom) correspond to nodes clustered with $spn$ values of 1, 8, 32 and 128. The left column shows the spatial distribution of nodes, the center column the corresponding 2D Fourier power spectrum, and the right column the radially averaged power spectrum.

The $spn=1$ case corresponds to a uniformly random distribution of nodes. The corresponding 2D-Fourier power spectrum exhibits an even frequency content typical of white noise. As the $spn$ parameter increases, nodes tend to distribute more evenly in space, and the corresponding 2D-Fourier spectrum indicates that meshes remain isotropic (no radial variation in the spectrum) after the clustering algorithm. Due to this angular invariance, all spectral information can be exhamined through the radially averaged power spectrum. As it can be observed from the right column on Figure \ref{fig:KMeansSpectrum}, the clustering algorithm \emph{filters} the lower frequencies in the Fourier space, resulting in a typical blue noise power spectrum. That is, the lower the value of $spn$ the closer the tendency to generate to a purely white noise mesh, where as $spn$ increases the mesh approaches the behavior of a blue-noise mesh, which preserves isotropy and randomness while filtering low frequencies resulting in more even spacing between nodes.

\begin{figure}
\begin{center}
{\small
$\text{spn}=1$ \\(uniformly random distribution)}\\
\subfloat{\includegraphics[height=0.3\textwidth]{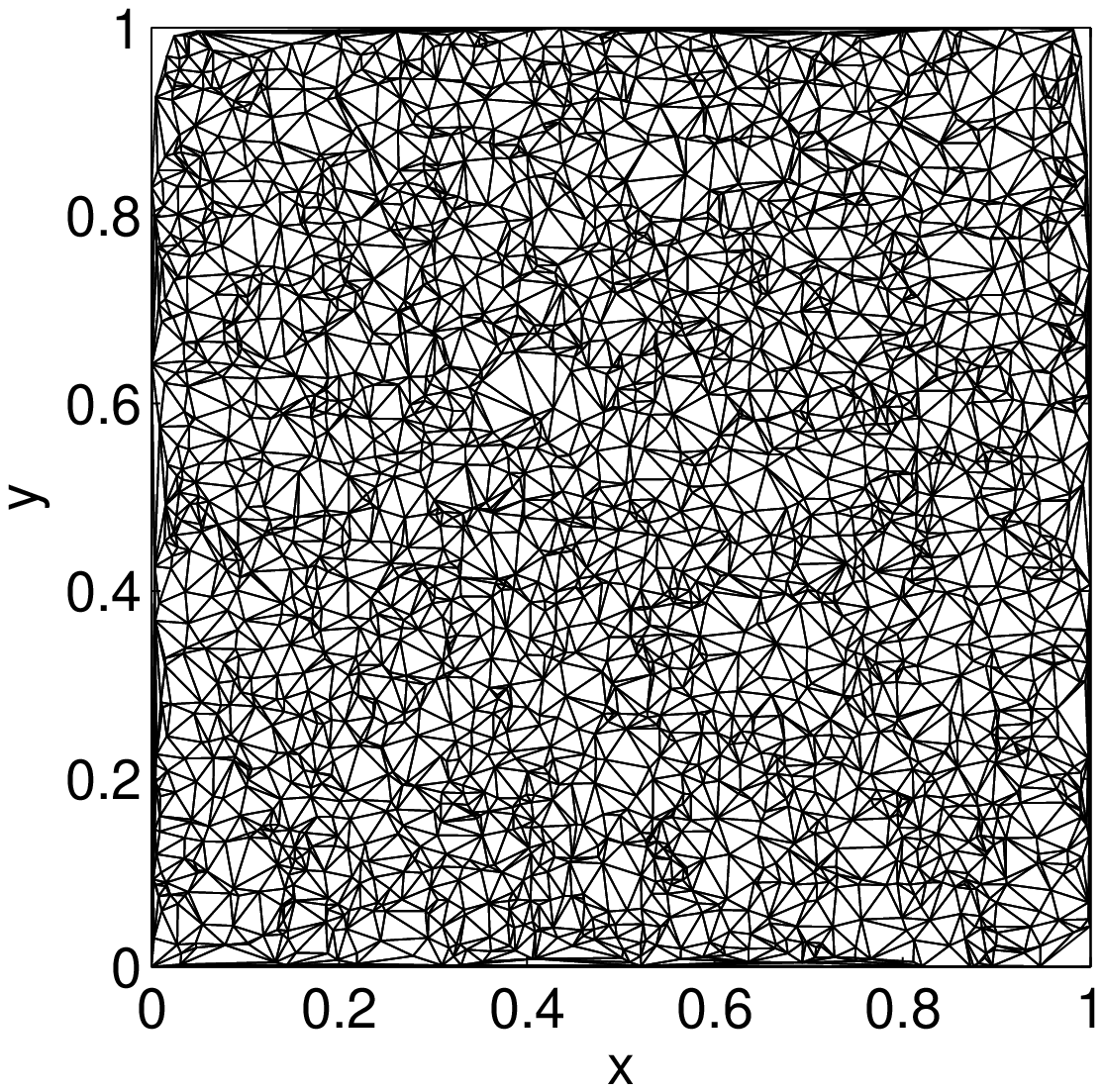}}
\subfloat{\includegraphics[height=0.3\textwidth]{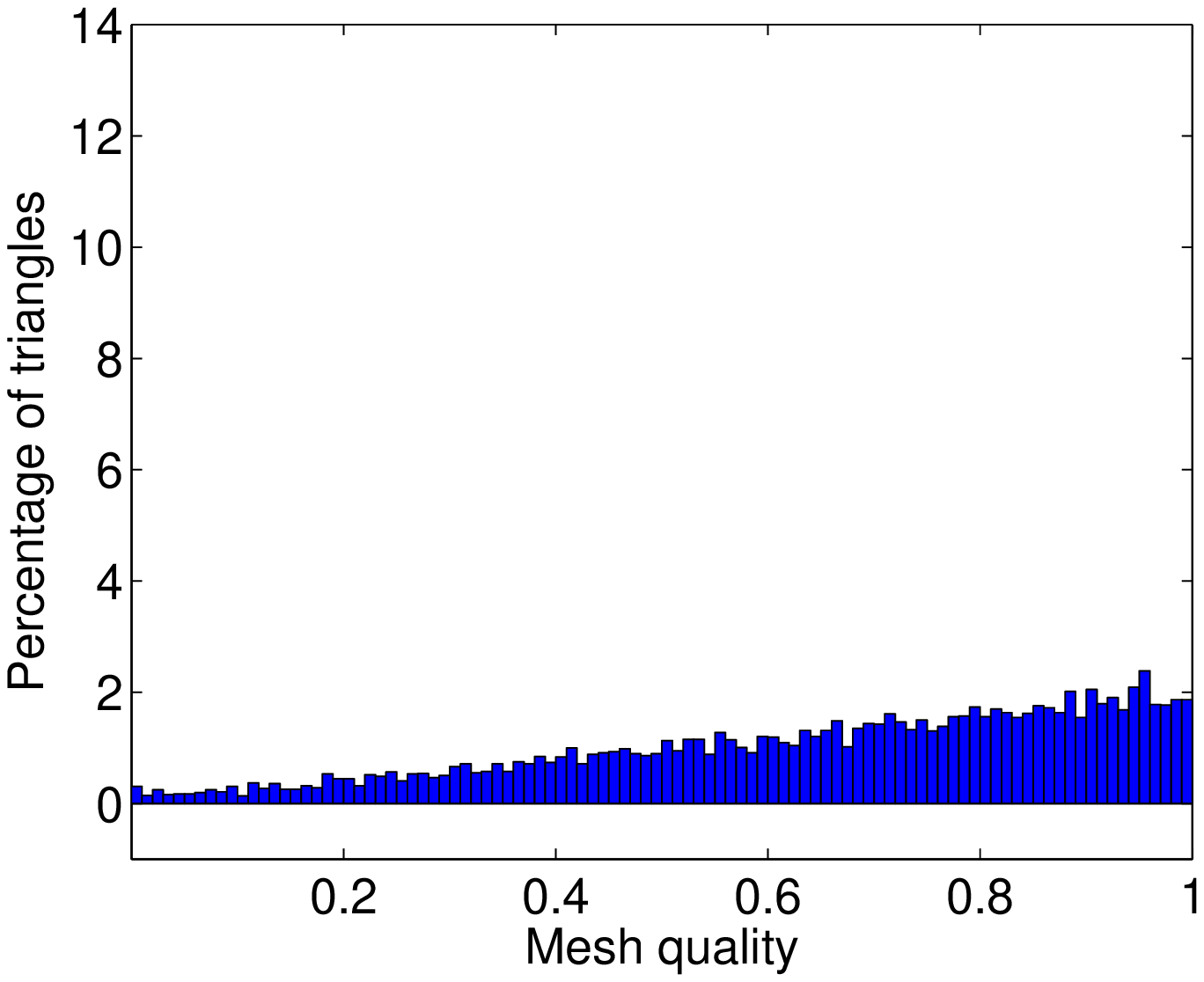}}\\
{\small
$\text{spn}=8$} \\
\subfloat{\includegraphics[height=0.3\textwidth]{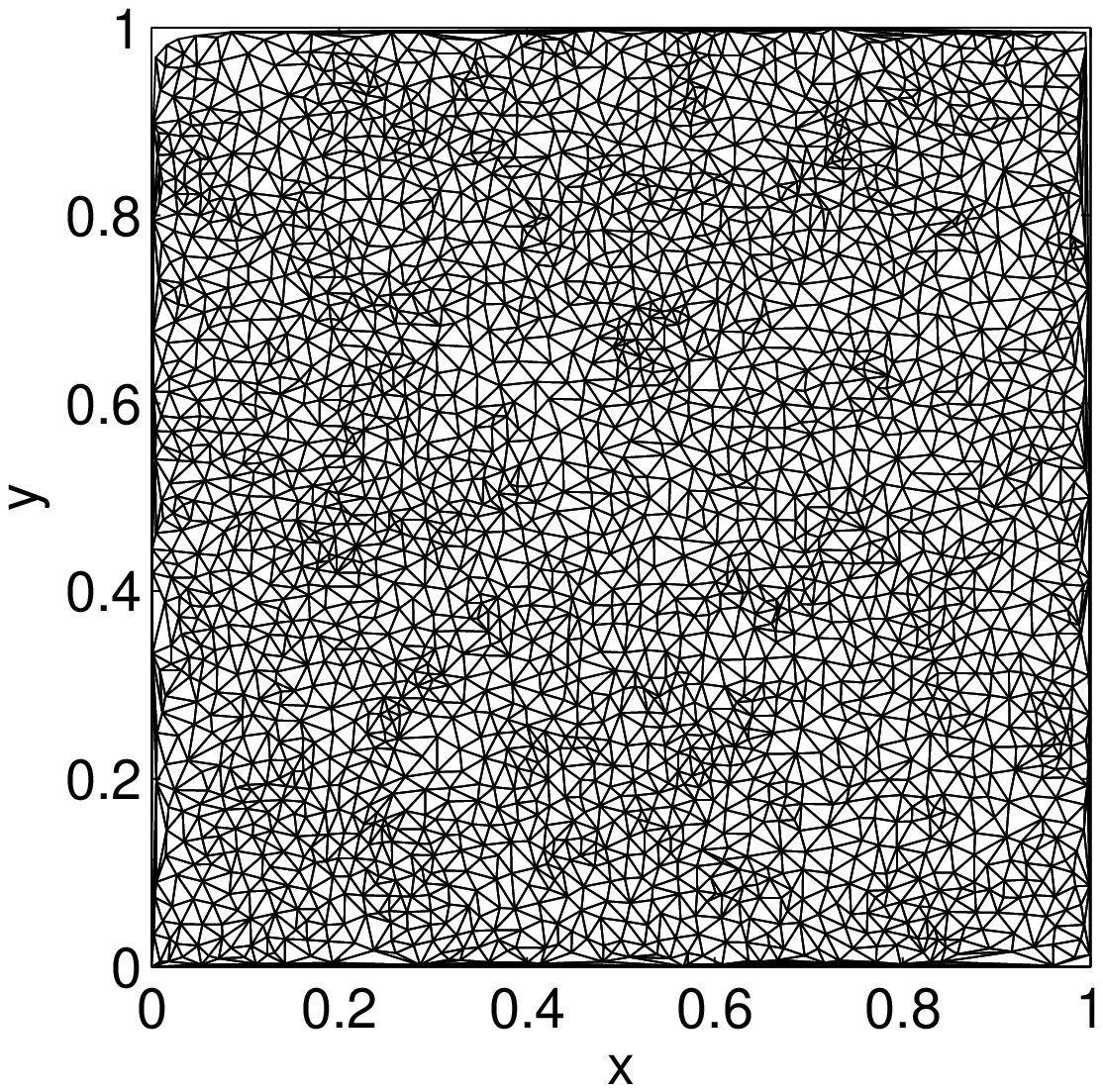}}
\subfloat{\includegraphics[height=0.3\textwidth]{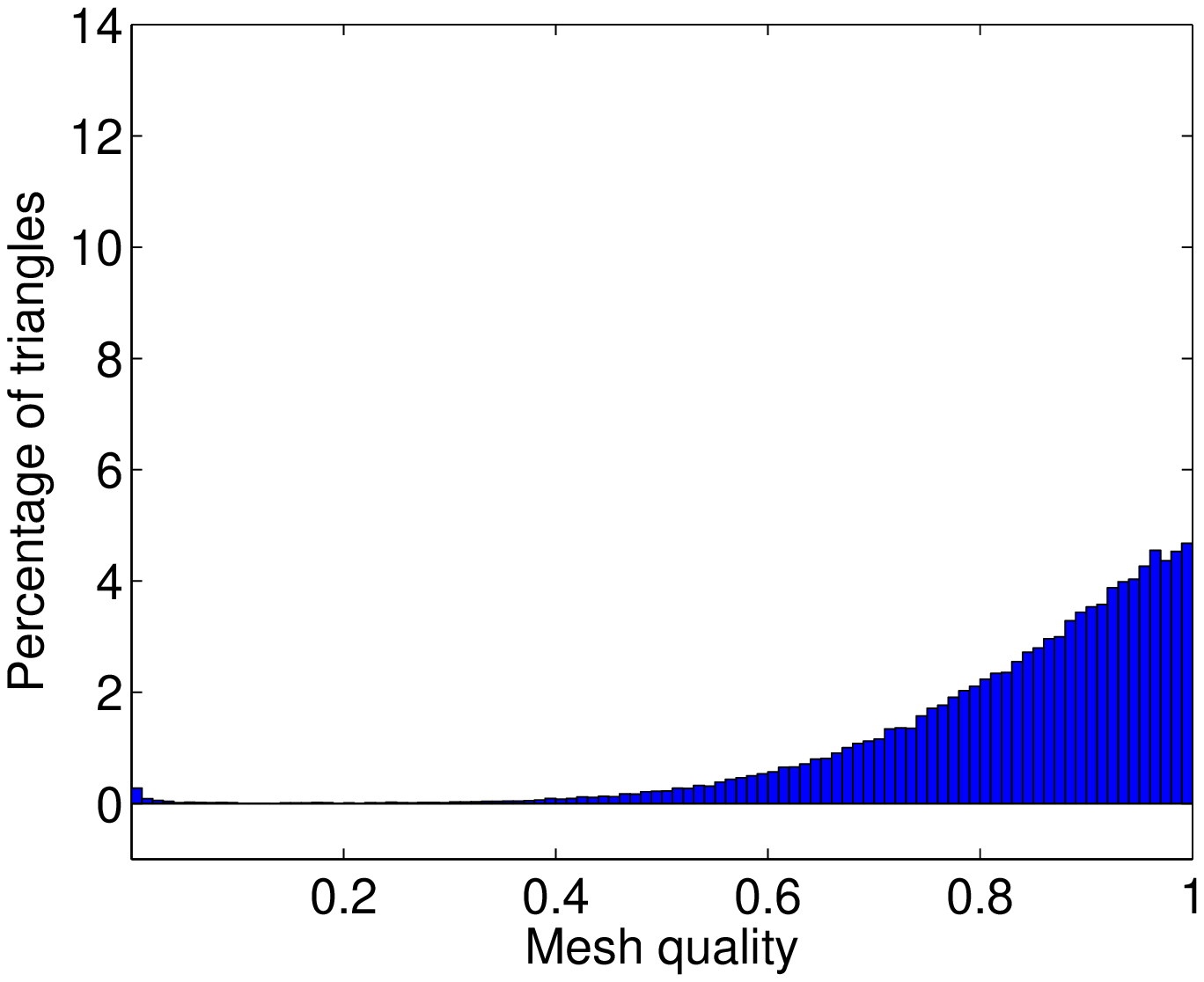}}\\
{\small
$\text{spn}=32$} \\
\subfloat{\includegraphics[height=0.3\textwidth]{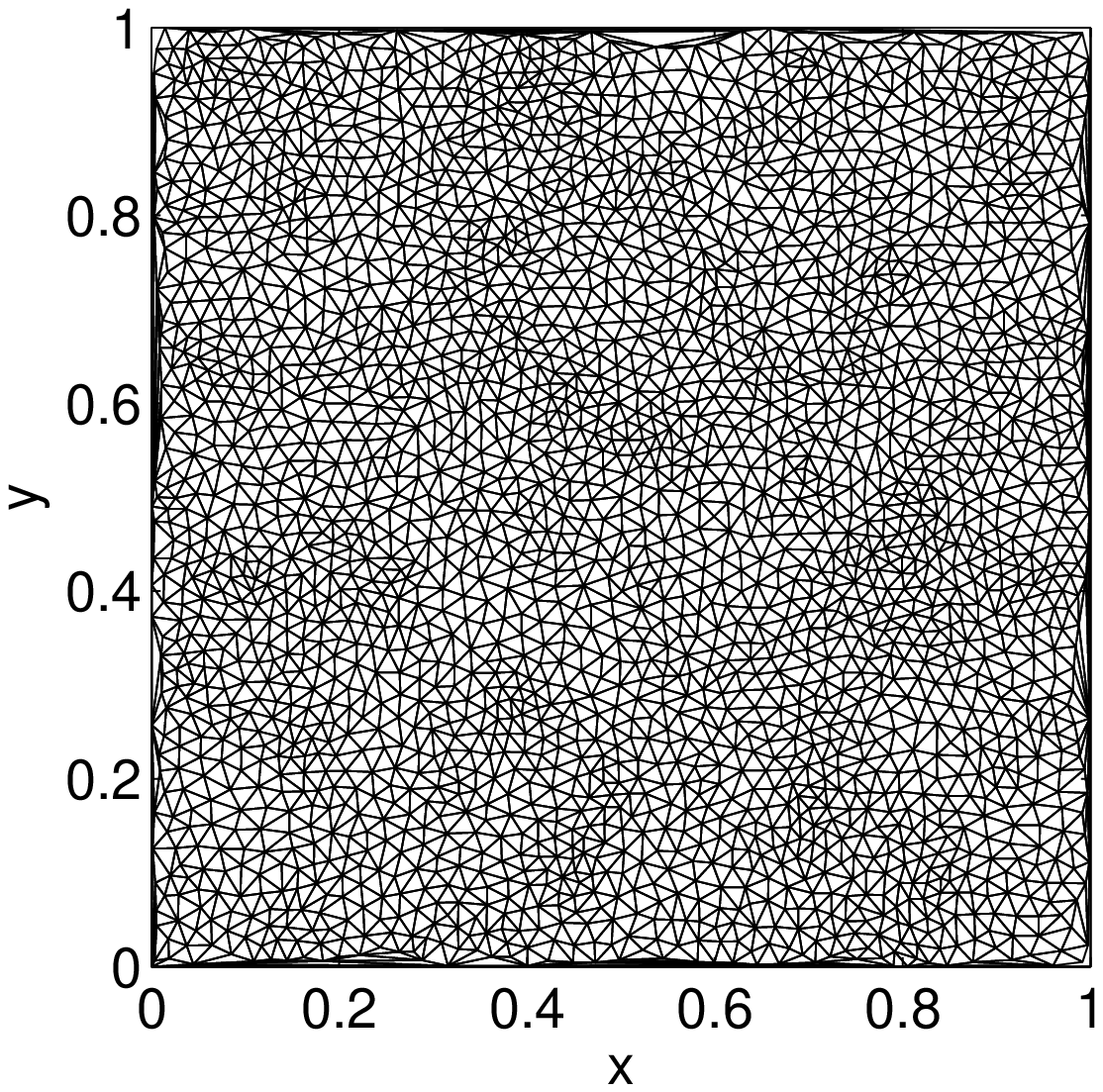}}
\subfloat{\includegraphics[height=0.3\textwidth]{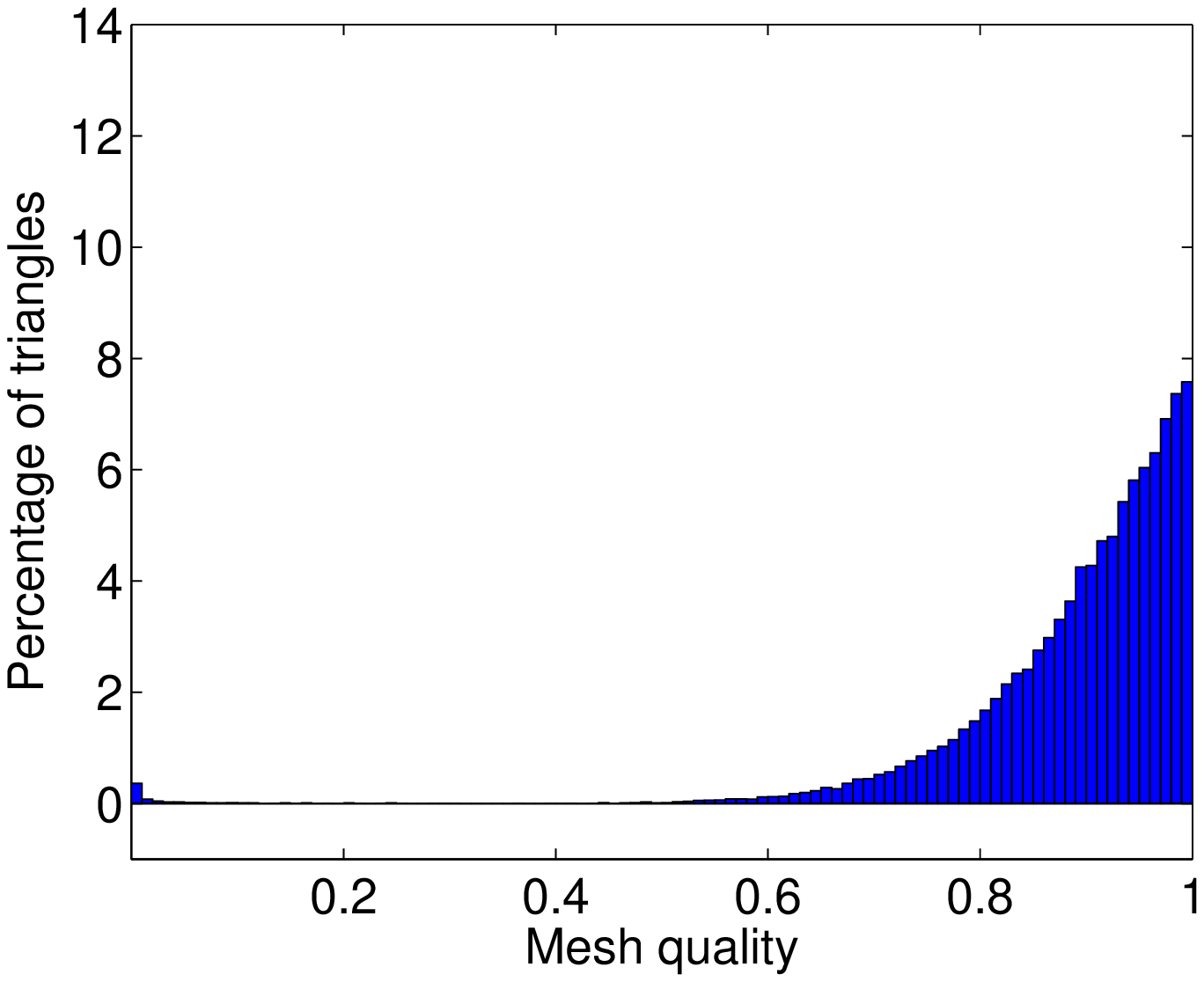}}\\
{\small
$\text{spn}=128$} \\
\subfloat{\includegraphics[height=0.3\textwidth]{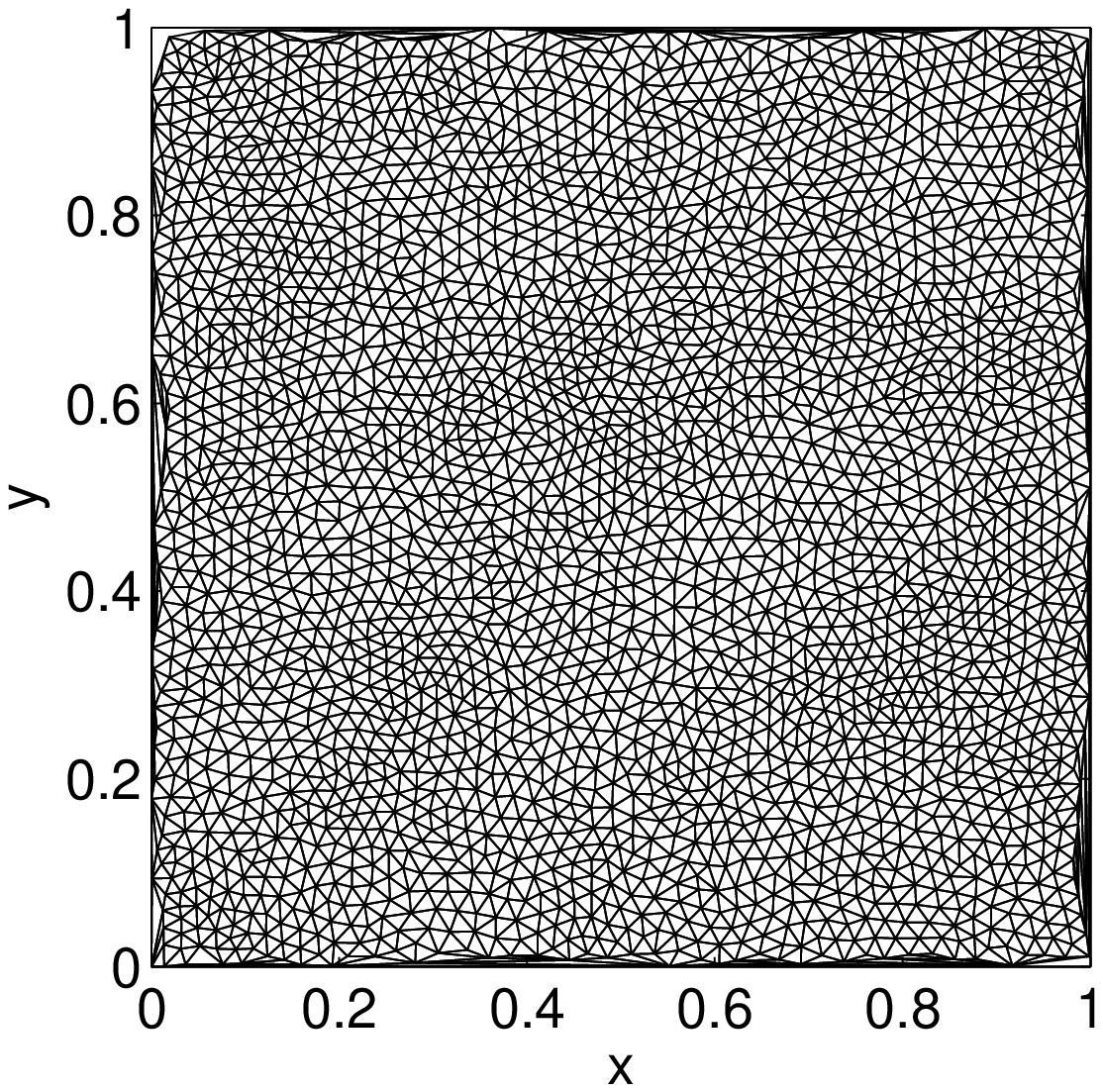}}
\subfloat{\includegraphics[height=0.3\textwidth]{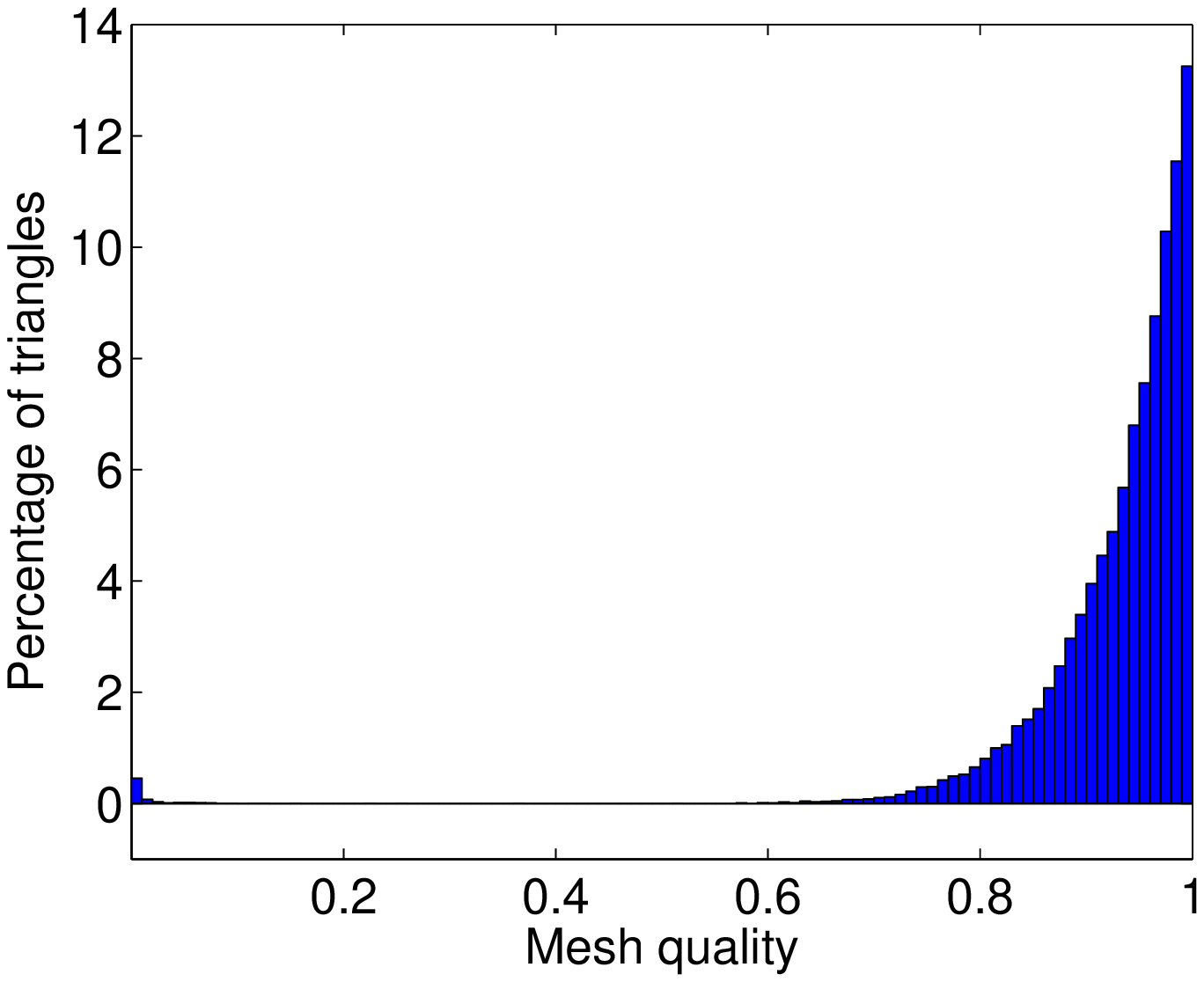}}\\
\end{center}
\caption{\small Influence of the smoothing parameter $spn$ on mesh quality. Mesh (left column) and corresponding mesh quality distribution (right column).}\label{fig:KMeansQuality}
\end{figure}

Figure \ref{fig:KMeansQuality} shows the effect of the clustering parameter $spn$ on the mesh quality distribution. The Figure is organized in a matrix layout where rows (from top to bottom) correspond to nodes clustered with $spn$ values of 1, 8, 32 and 128. The left column shows the obtained mesh, and the right column the distribution of the mesh quality parameter $q=2r/R$ within the mesh, where $r$ is the inradius and $R$ is the circumradius of a triangle in the mesh. The chosen mesh quality parameter can take values between 0 and 1, 0 indicating a sliver (or zero-volume triangle) and 1 indicating an equilateral triangle. Clearly, as the value of $spn$ increases the corresponding mesh quality distribution improves drastically. As a result, \emph{K-means meshes with higher value of the cluster parameter $spn$ tend to contain a larger proportion of nearly equilateral triangles}.

\begin{figure}
\begin{center}
\subfloat[]{\includegraphics[height=0.39\textwidth]{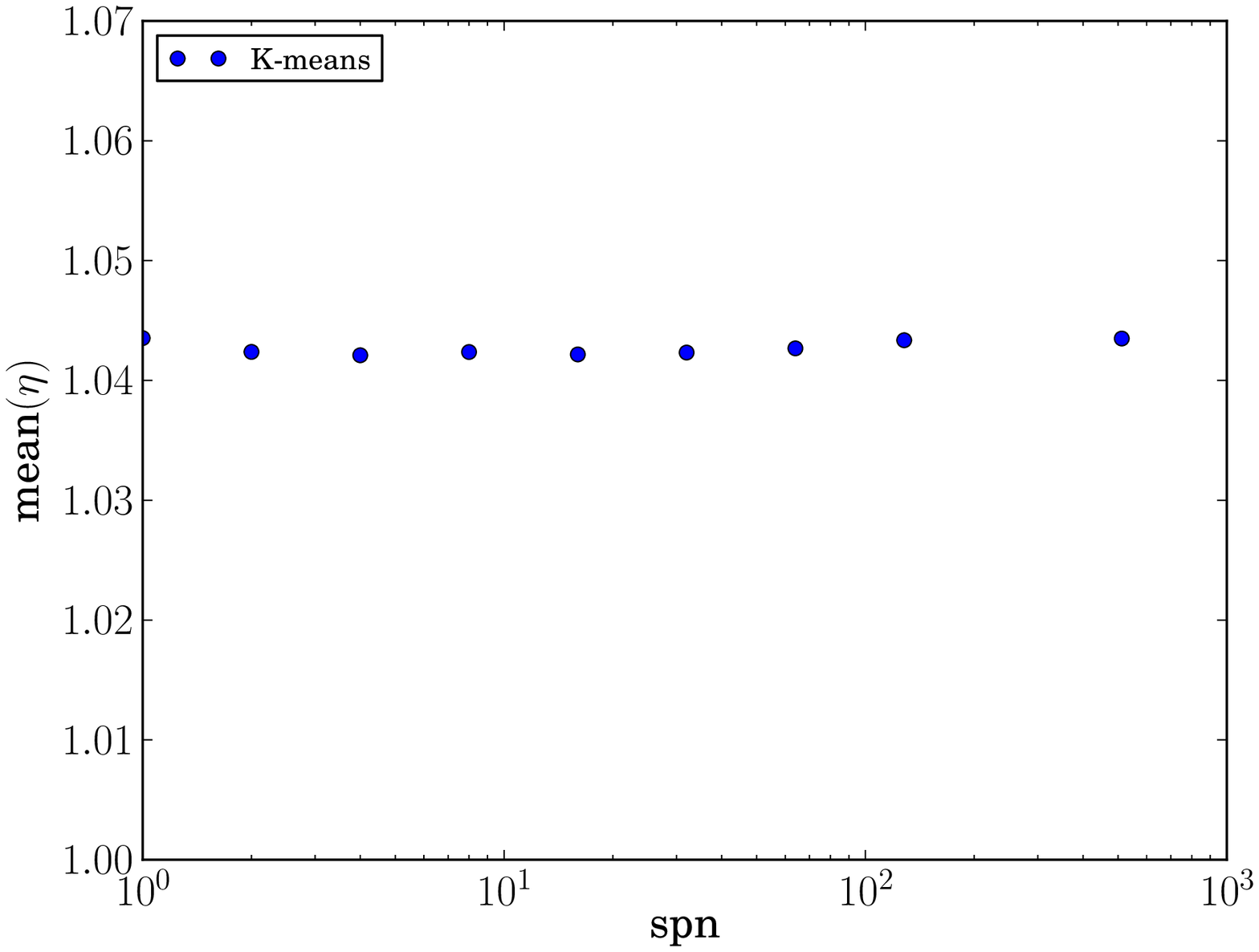}}
\subfloat[]{\includegraphics[height=0.39\textwidth]{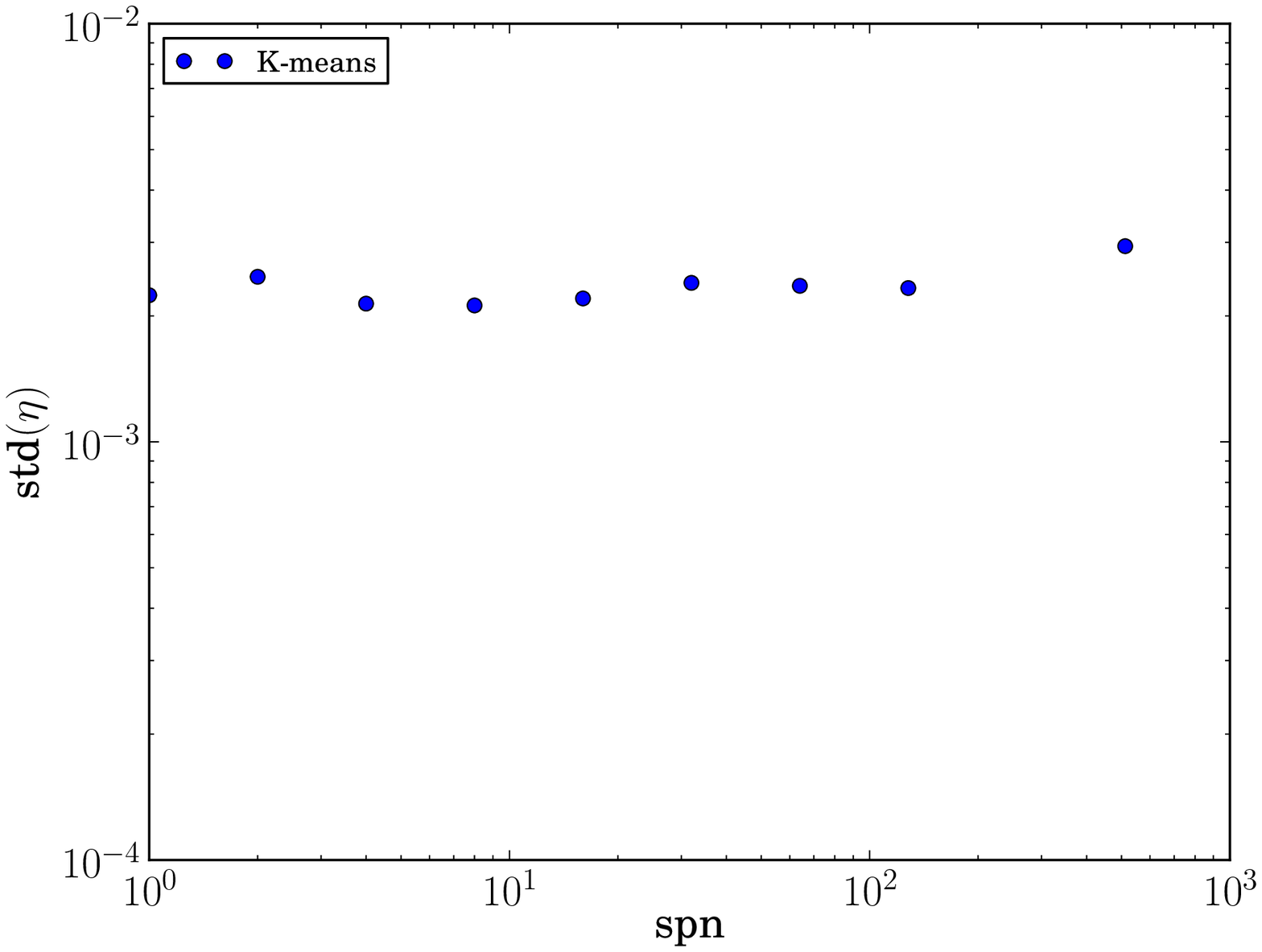}}\\
\end{center}
\caption{\small Influence of the smoothing parameter $spn$ on (a) $\mathrm{mean}(\eta)$, and (b) $\mathrm{std}(\eta)$ for a K-means mesh.}\label{fig:spnKMeans}
\end{figure}

To conclude our study of K-means meshes, we analyze the effect of the clustering parameter $spn$ on the metrics relevant to our original problem: the path deviation ratio. Figure \ref{fig:spnKMeans}a depicts the value of the mean value of the path deviation ratio as a function of $spn$ for a mesh with non-dimensional mesh size $\lambda\approx 1/400$. As it can be observed, an increment on the value of $spn$ initially produces almost no change in the mean value of the path deviation ratio. Additionally, Figure \ref{fig:spnKMeans}b indicates that the standard deviation of the path deviation ratio is not much affected by the value of $spn$. That is, the smoothing process performed through the clustering algorithm improves the mesh quality while not affecting the behavior of path deviation ratio.

In summary, K-means meshes provide a good alternative to 4k meshes with NP and ES operators, as they provide good quality triangles and slightly higher values for the mean value of the path deviation ratio, while exhibiting a perfectly isotropic behavior.

\subsection{Reducing mesh induced toughness through conjugate-directions meshes}
The good isotropic behavior of the K-means meshes could be attributed to the ability of the mesh, at any node, to provide a set of random directions for the crack to continue its propagation towards a neighboring node. Similarly, the mean value of the path deviation ratio can be reduced by enriching the space of possible directions available at a point for a crack to propagate to the next point in the mesh, as done by the ES operator in 4k meshes. Towards this end, we introduce a new type of mesh, which we term conjugate-directions mesh. This new family of meshes preserves isotropy by using as starting point an existing K-means mesh, and enriches the space of possible directions by introducing topological modifications to it. Conjugate-directions meshes are obtained as follows:
\begin{itemize}
\item A K-means mesh is generated
\item A barycentric subdivision is applied to the initial K-means mesh
\end{itemize}

An algorithm for performing a barycentric subdivision to simplicial meshes of any dimension is detailed in \cite{rimoli2011ijnme}, and explained in the subsequent paragraphs for completeness. Towards this end, we first describe (briefly) the concept of simplicial complex and subdivisions of a simplicial complex (triangulation). 

A cell complex is a collection of objects, or cells, to which a precise dimension can be assigned. At its fundamental level, a 3-dimensional object can be represented by a collection of 0-cells (vertices), 1-cells (edges), 2-cells (faces) and 3-cells (volumes). In the subsequent paragraphs, p-dimensional cells are denoted as $e_p$. In addition, $E_p(\mathcal S)$ represents the collection of p-dimensional cells in a complex $\mathcal S$. The dimension $\dim \mathcal S$ of the cell complex $\mathcal S$ is the largest dimension of any of its cells. Special classes of cell complexes are obtained when the cells are restricted to be of a certain type.

In geometry, a simplex is a generalization of the notion of a triangle or tetrahedron to arbitrary dimension. If all cells in a cell complex are {\sl simplices}, the cell complex is said to be a {\sl simplicial complex}. Simplicial complexes arise naturally as a result of the triangulation of solids. Specifically, a simplicial complex $\mathcal S$ in $\mathbb R^n$ is a collection of simplices in $\mathbb R^n$ such that (i) every face of a simplex of $\mathcal S$ is in $\mathcal S$, and (ii) the intersection of any two simplices of $\mathcal S$ is a face of each of them.

We recall that the simplex $\sigma$ spanned by a geometrically independent point set $v_0,...,v_p$ in $\mathbb R^n$ is
\begin{equation}
\sigma=\left\{ x \in \mathbb R^n \, : \quad \sum_{i=0}^p \lambda_i = 1, \quad \sum_{i=0}^p \lambda_i v_i = x, \quad \lambda_i \geq 0 \quad \forall i=0,1,...,p\right\}
\end{equation}
where the numbers $\lambda_i$ are the barycentric coordinates of $x$ with respect to $v_0,...,v_p$. The barycentric coordinates of a point are uniquely determined by it. The points $v_0,...,v_p$ are the vertices of the simplex $\sigma$ and $p$ its dimension. Any simplex generated by a subset of ${v_0,...,v_p}$ is a proper face of $\sigma$ and the union of all proper faces is the boundary of the simplex. The notation $e_\beta \prec e_\alpha$ signifies that $e_\beta$ is a face of $e_\alpha$.

Let $\mathcal S$ be a cell complex. A {\sl subdivision complex} of $\mathcal S$ is obtained by subdividing its cells into finer cells. More precisely, a complex $\mathcal{S}^*$ is said to be a subdivision of $\mathcal S$ if (i) every cell of $\mathcal{S}^*$ is contained in a cell of $\mathcal S$, and (ii) every cell of $\mathcal S$ is the union of finitely many cells of $\mathcal{S}^*$. These conditions particularly imply that the union of the cells of $\mathcal{S}^*$ equals the union of the cells of $\mathcal S$ and, hence, $|\mathcal{S}^*| =|\mathcal S|$ as sets. The way in which a subdivision complex $\mathcal{S}^*$ is nested within the supercomplex $\mathcal S$ may be described by means of an inclusion map $f:\mathcal{S}^* \rightarrow \mathcal S$. This map assigns to every cell $e_\beta$ of $\mathcal{S}^*$ the cell $e_\alpha$ of $\mathcal S$ that contains it.

{\sl Barycentric subdivision} generates a particular class of subdivision complexes. Thus, the barycentric subdivision of a simplicial complex is obtained by means of a uniform refinement. If $\sigma_p=[v_0,...,v_p]$ is a p-simplex of a simplicial complex $\mathcal S$, its barycenter is the point
\begin{equation}
\hat\sigma_p=\sum_{i=0}^{p}\frac{1}{p+1}v_i
\end{equation}
i.~e., the barycenter is the point in the interior of $\sigma_p$ that has equal barycentric coordinates. Given a simplicial complex $\mathcal S$ its barycentric subdivision $\mathcal{S}^*$ consists of all simplices $[\hat\sigma_0,...,\hat\sigma_p]$ such that $\hat\sigma_0 \succ ... \succ \hat\sigma_p$.  The details of the baycentric subdivision algorithm are depicted in Algorithm {\ref{alg:BarycentricSubdivision} for completeness. As an example, Figure \ref{fig:BarycentricSubdivision} shows the barycentric subdivision of a triangle.  It is worth noting that the barycentric subdivision of a triangle will produce 6 new triangles whose aspect ratio is worse than that of the the original one.

\begin{figure}
\begin{centering}
  \includegraphics[width=0.25\textwidth]{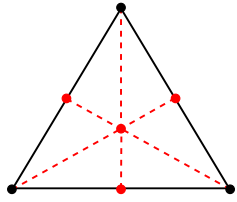}\\
  \caption{\small Barycentric subdivision of a triangle.}
  \label{fig:BarycentricSubdivision}
\end{centering}
\end{figure}

\begin{algorithm}
\KwIn{simplicial complex $\mathcal{S}$, $n=\dim{\mathcal{S}}$.}
\KwOut{barycentric complex $\mathcal{S}^*$.}
\tcc{Initialize barycentric complex $\mathcal{S}^*$ and inclusion map $f$}
\For{$(p=0,...,n)$}
{
    \For{$(e_p \in E_p(\mathcal S))$}
    {
        insert $v$ in $E_0(\mathcal{S}^*)$\;
        insert $(e_p, v)$ in $f$\;
    }
}
\tcc{Compute connectivity table $CT(\mathcal S)$}
\For{$(e_n \in E_n(\mathcal S))$}
{
    insert $[e_n]$ in $CT(\mathcal S)$\;
}
\For{$(p=n,...,1)$}
{
    \For{$([e_n,...,e_p] \in CT(\mathcal S))$}
    {
        \For{$(e_{p-1} \prec e_p)$}
        {
            append $e_{p-1}$ to $[e_n,...,e_p]$\;
        }
    }
}
\tcc{Compute connectivity table $CT(\mathcal{S}^*)$}
\For{$([e_n,...,e_0] \in CT(\mathcal S))$}
{
    insert $[f(e_n),...,f(e_0)]$ in $CT(\mathcal{S}^*)$\;
}
construct the simplicial complex $\mathcal{S}^*$ from $CT(\mathcal{S}^*)$\;
\KwRet{$\mathcal{S}^*$}
\caption{Barycentric subdivision.}
\label{alg:BarycentricSubdivision}
\end{algorithm}

An interesting feature of the barycentric subdivision, and the reason why we chose it to generate conjugate-directions meshes, is that \emph{in the limiting case of equilateral triangles, the barycentric subdivision of a triangle provides an orthogonal direction for each direction provided by the edges of the original triangulation}. This can be clearly observed in Figure \ref{fig:BarycentricSubdivision}, where there is a newly introduced dashed line perpendicular to each edge of the original triangle. In this way, for K-means meshes with high value of $spn$ (for which we already illustrated their higher content of equilateral triangles), their barycentric subdivision enriches them with a set of directions that are almost orthogonal to the existing edges. Figure \ref{fig:Barycentric} shows a typical conjugate-directions mesh obtained from a K-means mesh with $spn=128$ and $n=512$.

\begin{figure}
\begin{center}
\subfloat[]{\includegraphics[height=0.45\textwidth]{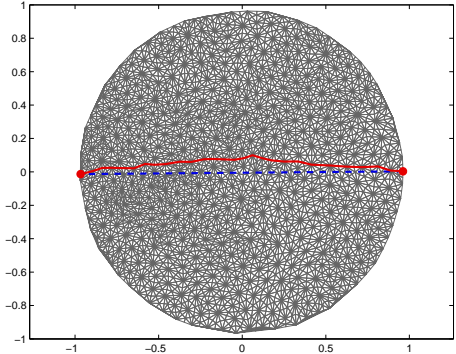}}\\
\end{center}
\caption{\small Typical conjugate-directions mesh obtained from a K-means mesh with $spn=128$ and $n=512$.}\label{fig:Barycentric}
\end{figure}

Figure \ref{fig:ComparisonAll} shows the values of the relative error $\epsilon$ vs direction for a 4k mesh with NP and ES, a K-means mesh, and a conjugate-directions mesh. All meshes are comparable in terms of their non-dimensional size $\lambda\approx 1/250$. The conjugate-directions mesh is not only isotropic, but the mean value of the error is significantly lower ($\approx 0.018$) than that observed for the other meshes ($\approx 0.04$).

\begin{figure}
\begin{center}
\includegraphics[height=0.48\textwidth]{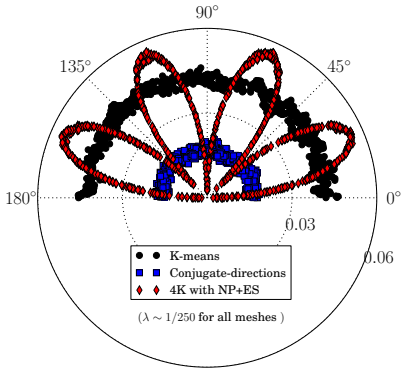}
\end{center}
\caption{\small Relative error ($\epsilon = \eta-1$) vs mesh direction for a 4k mesh with NP and ES, a K-means mesh, and a conjugate-directions mesh. All non-dimensional mesh sizes correspond to $\lambda\approx1/250$.}\label{fig:ComparisonAll}
\end{figure}

It is worth noting that, since conjugate-directions meshes are derived from K-means meshes, the $spn$ parameter of the seed mesh will play a role in the performance of the newly generated one. Figure \ref{fig:spnBary} shows the effect of the $spn$ parameter on the path deviation ratio for a conjugate-directions mesh with $\lambda\approx1/400$. As shown in \ref{fig:spnBary}a, an increment of $spn$ produces a consistent reduction on the mean value of the path deviation ratio. For $spn=512$ the mean value of the path deviation ratio reaches a value $\text{\bf mean}(\eta)\approx 1.015$, corresponding to a mean value of the relative error $\text{\bf mean}(\epsilon)\approx 0.015$ which is roughly half of that observed for the barycentric mesh obtained from a random ($spn=1$) seed mesh. Additionally, Figure \ref{fig:spnBary}b indicates that the standard deviation of the path deviation ratio decreases as $spn$ increases. This implies that, when a K-means mesh with high value of $spn$ is adopted as a generator of a conjugate-directions mesh, the resulting triangulation produces a more compact distribution of the path deviation ratio.

In summary, conjugate-directions meshes provide an even better alternative to 4k meshes with NP and ES, as they provide significantly better values for the mean value of the path deviation ratio while being perfectly isotropic. It is worth noting that the original seed mesh must be of good quality since the barycentric subdivision stage in the generation of a conjugate-directions mesh produces a deterioration in the quality of the triangulation.

\begin{figure}
\begin{center}
\subfloat[]{\includegraphics[height=0.39\textwidth]{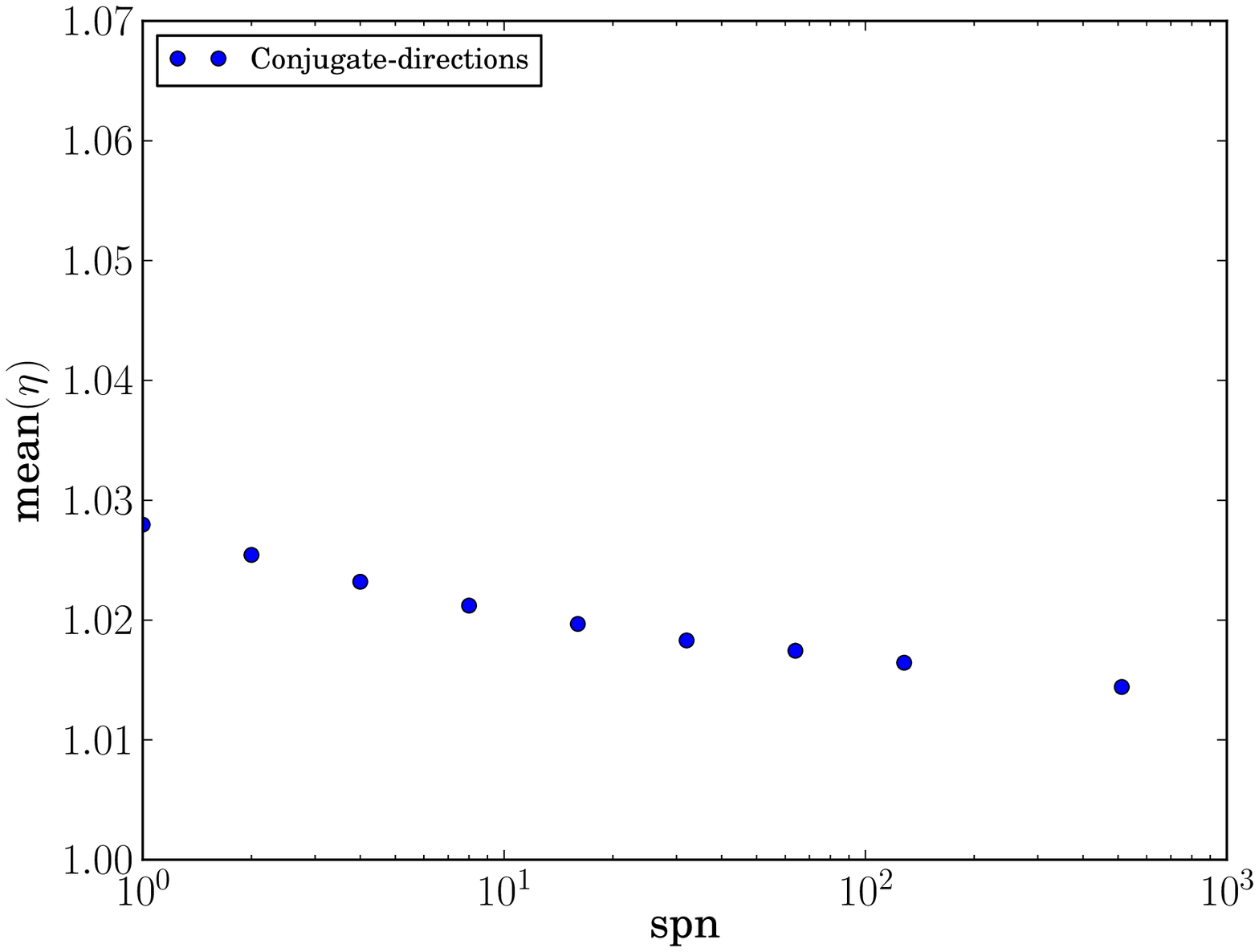}}
\subfloat[]{\includegraphics[height=0.39\textwidth]{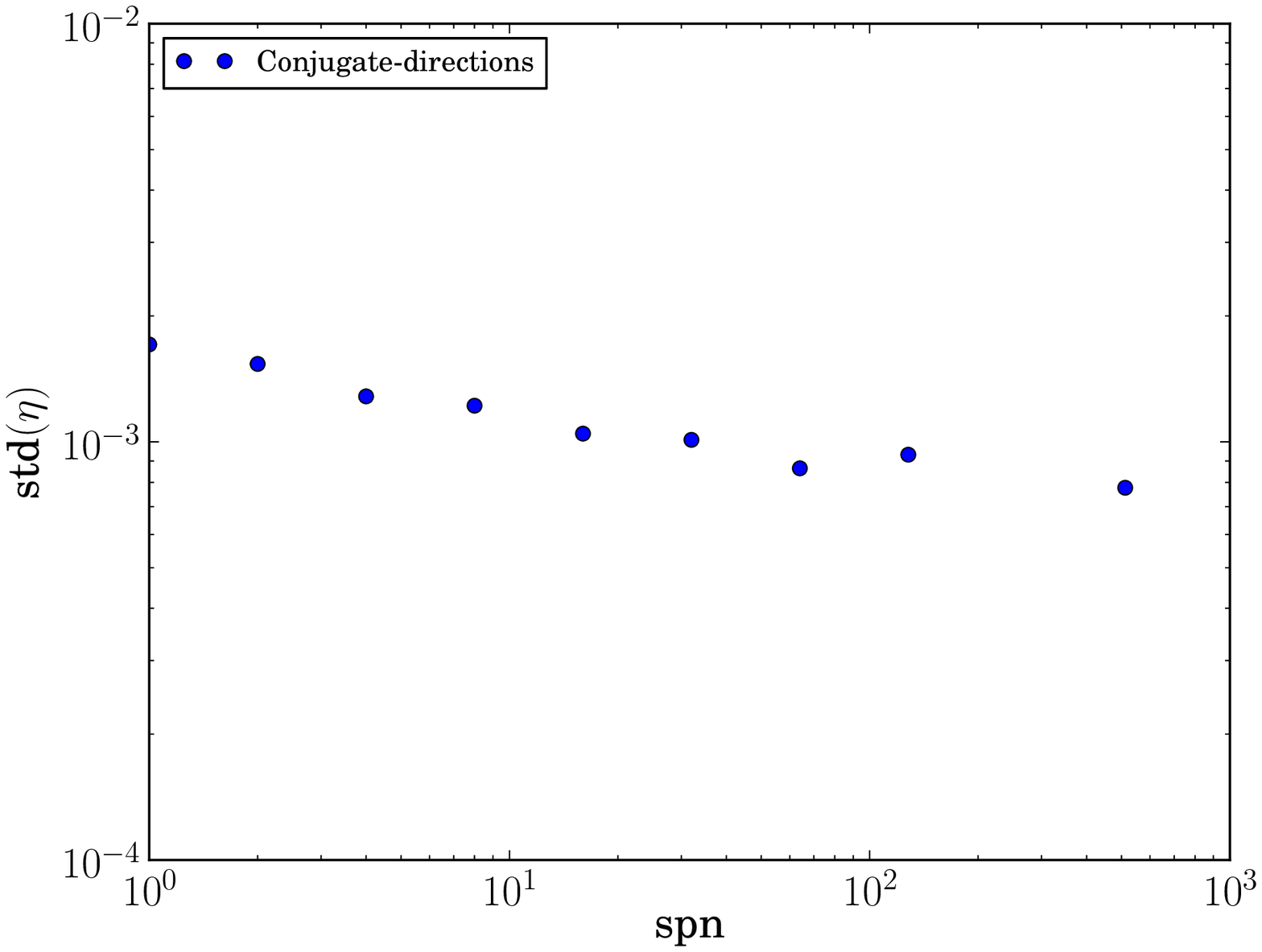}}\\
\end{center}
\caption{\small Influence of the smoothing parameter $spn$ on (a) $\mathrm{mean}(\eta)$, and (b) $\mathrm{std}(\eta)$ for a conjugate-directions mesh.}\label{fig:spnBary}
\end{figure}

\subsection{Convergence of K-means and conjugate-direction meshes}

To conclude this section, we perform a numerical convergence analysis of 4k, K-means and conjugate-direction meshes for the non-dimensional mesh size $\lambda$ tending to zero. For 4k and K-means meshes $\lambda$ ranges from $\approx 1/15$ to $\approx 1/360$. Since conjugate direction meshes were obtained from previously generated K-means meshes and the barycentric subdivision reduces the mesh size, $\lambda$ ranges form $\approx 1/43$ to $\approx 1/808$ for these meshes. All K-means and conjugate-directions meshes considered in this numerical study of convergence were generated with a clustering parameter $spn=128$. The nodal perturbation factor adopted for all 4k meshes is $\text{NP}_f=0.3$. 

Results show that convergence of K-means meshes, in the sense of the mean value of the path deviation ratio, is similar to that of 4k meshes with NP as the non-dimensional mesh size is reduced, see Figure \ref{fig:convergence}a. Overall, 4k meshes with NP and ES exhibit lower mean values of the path deviation ratio within the mesh size range under consideration. This difference seems to decrease as the meshes are refined. Numerical evidence also shows that the mean value of the path deviation ratio tends to saturate around 1.04 for 4k with NP and K-means meshes, and around 1.036 for 4k mesh with NP and ES. That is, these three types of mesh have an \emph{intrinsic roughness} that prevents them from representing straight lines with errors below 4\% or 3.6\% respectively no matter how refined they are.

On the other hand, the standard deviation of the path deviation ratio for K-means meshes decreases at a fastest rate when compared to both types of 4k meshes, see Figure \ref{fig:convergence}b. This indicates that even though both kinds of mesh exhibit similar mean values of the path deviation ratio for all mesh sizes, K-means meshes show less dispersion as the mesh is refined, while 4k meshes saturate very quickly. The saturation in the standard deviation observed on 4k meshes is a consequence the fact that the mesh-induced anisotropy does not disappear as the meshes are refined, see Figure \ref{fig:convergencepolar}a.

In addition, the same numerical experiment shows no indication of saturation for conjugate-directions meshes in the studied range of non-dimensional mesh sizes. Furthermore, the mean value of the path deviation ratio is significantly smaller (on the order of 1/3 to 1/2 in relative error terms) for conjugate-directions meshes when compared to the other meshes for all non-dimensional mesh sizes considered in this study. The standard deviation of the path deviation ratio seems to converge at a slower rate for this kind of mesh when compared to K-means meshes, but at a faster rate when compared to 4k meshes. It is worth noting, however, that it seems that the convergence rate on the standard deviation of the path deviation ratio tends to accelerate as the mesh is refined. This decreasing tendency on the standard deviation of the path deviation ratio indicates that the path deviation tends to concentrate around the mean value for conjugate-directions meshes as they are refined, see Figure \ref{fig:convergencepolar}b.

\begin{figure}
\begin{center}
\subfloat[]{\includegraphics[height=0.39\textwidth]{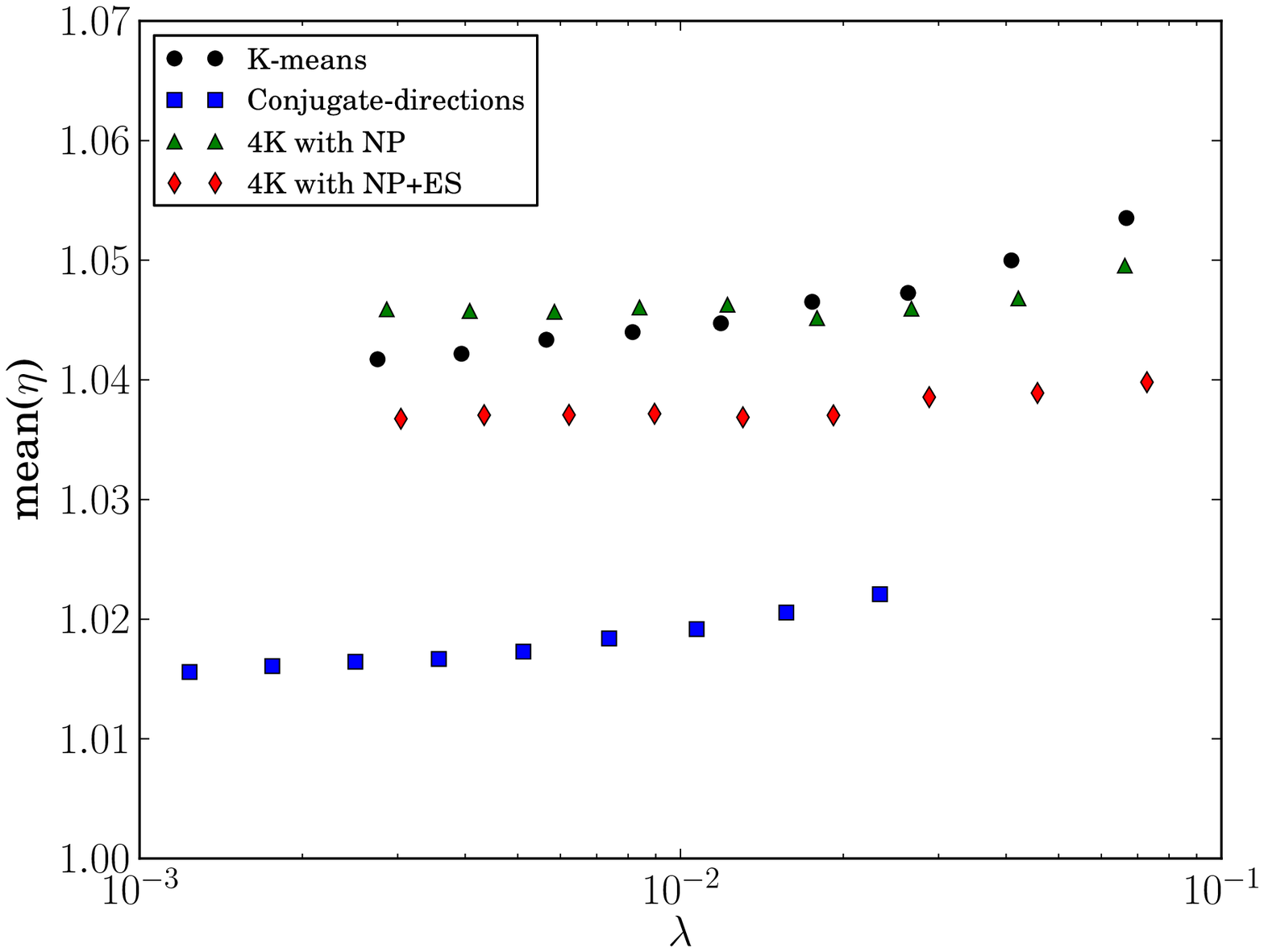}}
\subfloat[]{\includegraphics[height=0.39\textwidth]{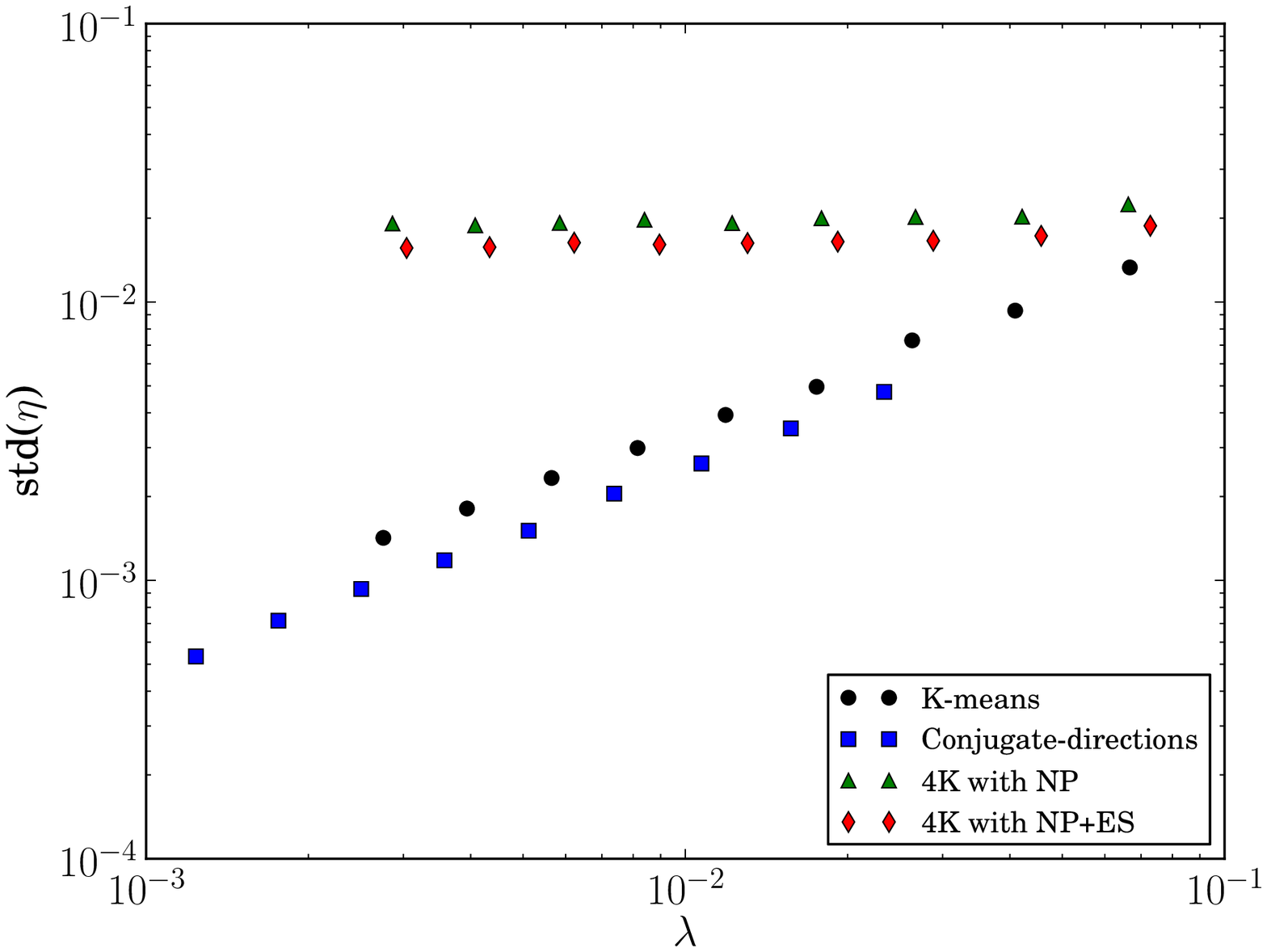}}\\
\end{center}
\caption{\small Numerical convergence analysis of 4k, K-means, and congugate-directions meshes. Mean value of path deviation ratio (a) and standard deviation of path deviation ratio (b) as a function of the non-dimensional mesh size.}\label{fig:convergence}
\end{figure}

\begin{figure}
\begin{center}
\subfloat[]{\includegraphics[height=0.48\textwidth]{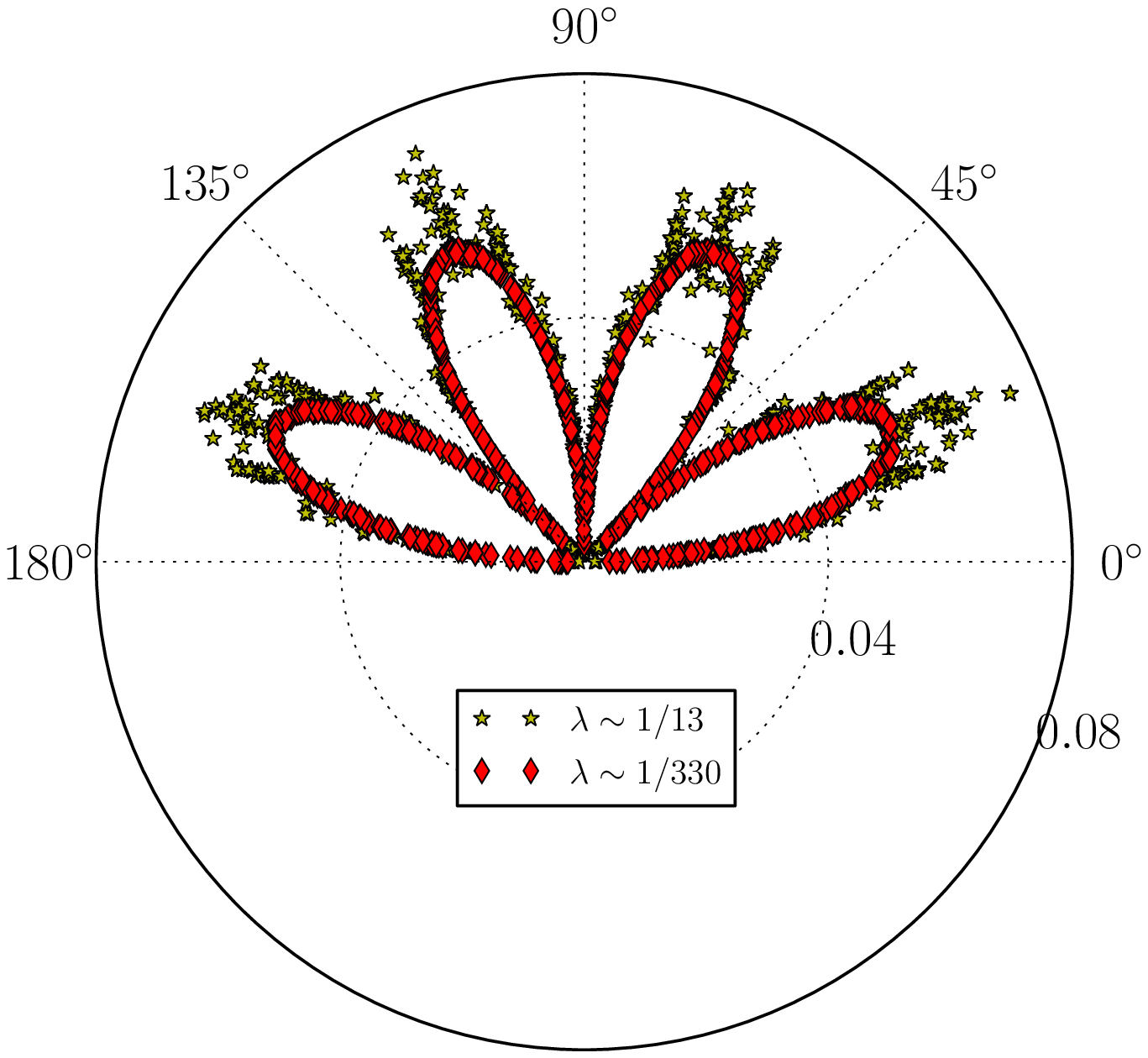}}
\subfloat[]{\includegraphics[height=0.48\textwidth]{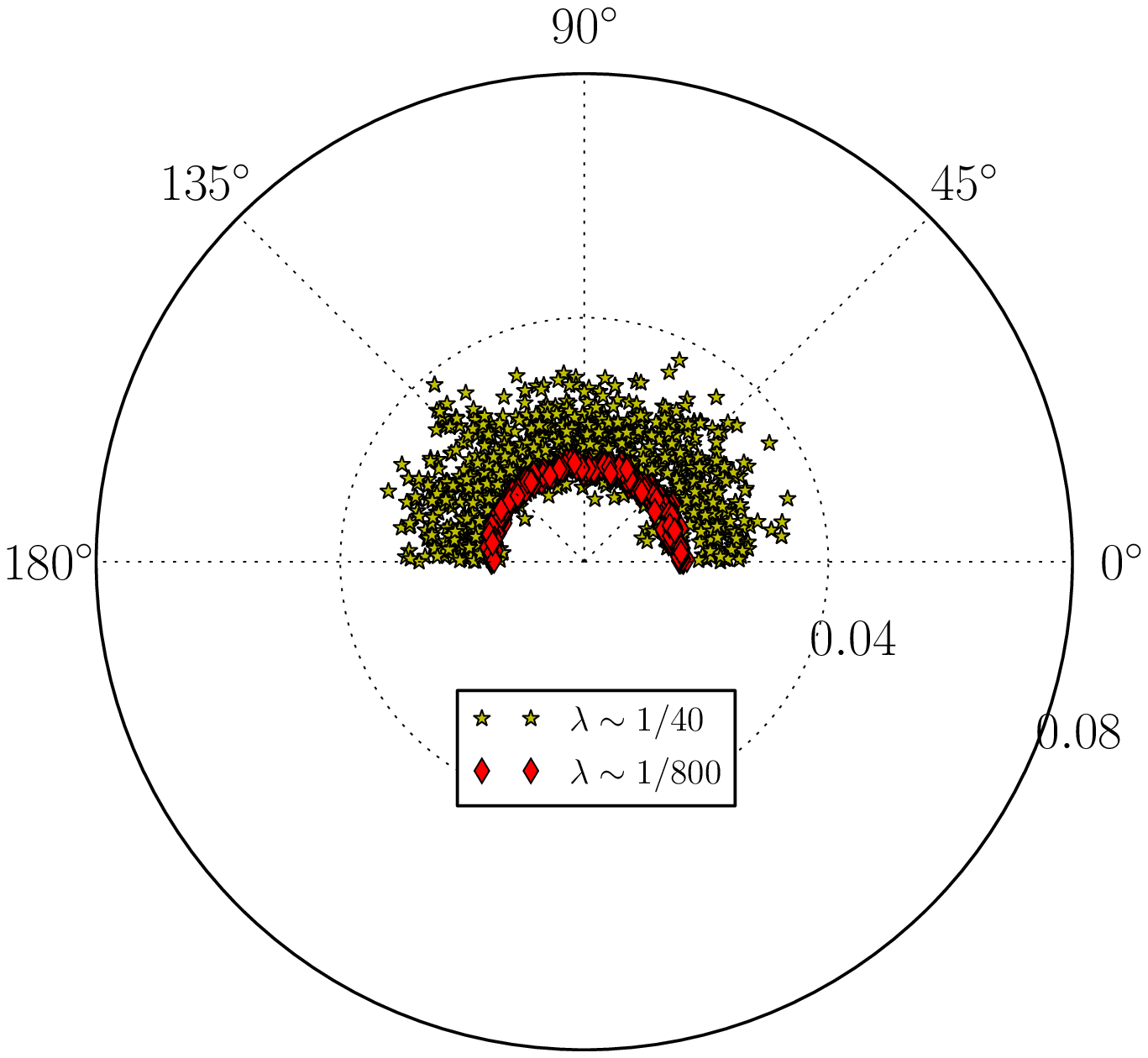}}\\
\end{center}
\caption{\small Polar behavior of the relative error ($\epsilon=\eta-1$) before and after mesh refinement for (a) 4K mesh with ES and NP, and (b) for conjugate-directions mesh.}\label{fig:convergencepolar}
\end{figure}

\section{Summary and final remarks}\label{sec:SUM}
Through a curve rectification reasoning, we showed that in order for a mesh to be able to represent an arbitrary crack through its edges, it should able to at least represent a straight segment in any arbitrary direction. We then illustrated how the introduction of a discretization produces mesh-induced anisotropy and mesh-induced toughness when dealing with cohesive element modeling of arbitrary cracks. Subsequently, we analyzed those effects through the introduction of polar plots of the relative error (or path deviation ratio), and showed that randomness plays a crucial role in the reduction of the former effect.

We observed that 4k meshes with NP and ES provide a better alternative to plain 4k meshes in the sense that they reduce the mesh-induced toughness. However, when looking at the full polar plot of the path deviation ratio versus mesh direction, it becomes evident that mesh-induced anisotropy, while reduced, still persists. That is, regardless of the type of 4k mesh (with or without NP and/or ES), there is always an associated anisotropy to it. In addition, this effect seemingly does not disappear as the mesh is refined.

On the other hand, K-means meshes were shown to be a good compromise between isotropy and good element quality. They are isotropic thus not providing preferred crack propagation directions for any mesh size. However, K-means meshes saturate, in the sense of the mean value of the path deviation ratio, as they are refined. This saturation results in the introduction of mesh-induced toughness even in the limiting case of non-dimensional mesh size tending to zero. A good feature of K-means meshes is that they do not seem to saturate in the sense of the standard deviation of the path deviation ratio, which implies that the error concentrates around its mean as the mesh is refined. This would allow to adopt K-means meshes in cohesive element modeling as long as the proper correction term is introduced in the toughness of the material.

We then proposed a new type of mesh, termed conjugate-directions mesh, which significantly reduces the undesired mesh-dependent effects for meshes of practical size. Our approach combines the use of barycentric subdivision with K-means meshes to provide a greater distribution of directions for failure surfaces to propagate. Numerical evidence suggest that, due to the exhibited isotropy and considerably lower mesh-induced toughness effect, conjugate-direction meshes could be good candidates for cohesive element analysis of crack propagation problems where crack paths are not known a priori. It is worth noting that the proposed method is based on: (i) generation of k-means distribution of nodes, (ii) a triangulation of those nodes, and (iii) a barycentric subdivision of the triangulation.  Point (i) is known to be able to handle graded node distribution and complex geometries in any dimension, point (ii) can be performed on any node distribution obtained through point (i), and point (iii) is topological thus well suited for any geometry in any dimension. That is, even though we focus on simple geometries for analysis purposes, our approach is applicable without need for further extension to complex geometries and graded meshes in arbitrary dimensions. However, the applicability of results obtained in 2D to the 3D case deserves further investigation.

Being local mesh refinement arguably the most widely adopted mesh adaption technique in the computational fracture mechanics community when dealing with cohesive element formulations, our observations of the convergence behavior of the path deviation ratio could have profound implications on the interpretation of mesh refinement algorithms. Even though mesh refinement schemes could improve the resolution of stress fields near crack tips, our research shows that \emph{at some point, depending on the meshing technique adopted, there could be absolutely no gain from mesh refinement in the sense of the energy dissipated by the crack no matter how much these meshes are refined}. Thus, we believe that to better understand the accuracy of mesh refinement schemes for cohesive element models, the relative impact on the accuracy of the solution of (i) the level of resolution of the stress field, and (ii) the correctness of the energy dissipated through the discrete cracks should be investigated.

To conclude, our hope is that by designing meshes to mitigate the effects of mesh geometry and topology, we can invoke adaptive insertion along pre-defined element boundaries to achieve arbitrary crack propagation. The great promise of this research effort is that arbitrary crack propagation could be achieved through mesh design (pre-processing).

\section*{Acknowledgments}

The author would like to acknowledge partial funding from Sandia National Laboratories under PO 1188989. Sandia National Laboratories is a multi-program laboratory managed and operated by Sandia Corporation, a wholly owned subsidiary of Lockheed Martin Corporation, for the U.S. Department of Energy’s National Nuclear Security Administration under contract DE-AC04-94AL85000. The authors would like to thank Dr. Alejandro Mota, Dr. James W. Foulk III, and Dr. Jakob Ostien from Sandia National Laboratories for their valuable discussions.

\end{document}